\documentclass[10pt,a4paper]{article}

\usepackage{authblk}
\usepackage{amssymb}
\usepackage{hyperref}
\usepackage{graphicx}
\usepackage{array}
\usepackage{siunitx}
\usepackage[usenames, dvipsnames]{color}
\usepackage{color}
\usepackage{subfig}
\usepackage{epsfig}
\usepackage{longtable}
\usepackage[utf8]{luainputenc}

\usepackage{epsfig}
\usepackage{epstopdf}
\usepackage{xspace}
\usepackage{multicol}

\usepackage{makecell}
\usepackage{cite}

\newcommand{\Smilei}{{\sc Smilei}\xspace}

\usepackage[margin=1.9cm]{geometry}
\usepackage{physics}

\usepackage{array}
\newcolumntype{L}{>{\centering\arraybackslash}m{3cm}}
\usepackage{multirow}

\hypersetup{%
   colorlinks=true,
   urlbordercolor={1 1 1},
   linkcolor={black},
   raiselinks=false,
citecolor=blue,
urlcolor=blue
}

\newcommand\blfootnote[1]{%
  \begingroup
  \renewcommand\thefootnote{}\footnote{#1}%
  \addtocounter{footnote}{-1}%
  \endgroup
}

\begin{document}

\title{\textbf{Numerical modeling of laser tunneling ionization in Particle in Cell Codes with a laser envelope model}}

\author{F. Massimo$^1$*, A. Beck$^1$, J. Derouillat$^2$, I. Zemzemi$^1$ and A. Specka$^{1}$}
\date{\small{$^1$ Laboratoire Leprince-Ringuet – École polytechnique, CNRS-IN2P3, Palaiseau 91128, France\\
$^2$ Maison de la Simulation, CEA, CNRS, Université Paris-Sud, UVSQ, Université Paris-Saclay, F-91191 Gif-sur-Yvette, France}}

\maketitle

\blfootnote{*Corresponding author. E-mail address: \href{mailto:massimo@llr.in2p3.fr}{massimo@llr.in2p3.fr}}

\begin{abstract}
The resources needed for Particle in Cell simulations of Laser Wakefield Acceleration can be greatly reduced in many cases of interest using an envelope model. However, the inclusion of tunneling ionization in this time averaged treatment of laser-plasma acceleration is not straightforward, since the statistical features of the electron beams obtained through ionization should ideally be reproduced without resolving the high frequency laser oscillations. In this context, an extension of an already known envelope ionization procedure is proposed, valid also for laser pulses with higher intensities, which consists in adding the initial longitudinal drift to the newly created electrons within the laser pulse ionizing the medium. The accuracy of the proposed procedure is shown with both linear and circular polarization in a simple benchmark where a nitrogen slab is ionized by a laser pulse, and in a more complex benchmark of laser plasma acceleration with ionization injection in the nonlinear regime. With this addition to the envelope ionization algorithm, the main phase space properties of the bunches injected in a plasma wakefield with ionization by a laser (charge, average energy, energy spread, rms sizes, normalized emittance) can be estimated with accuracy comparable to a non-envelope simulation with significantly reduced resources, even in cylindrical  geometry. Through this extended algorithm, preliminary studies of ionization injection in Laser Wakefield Acceleration can be easily carried out even on a laptop.
\end{abstract}

\section{Introduction}
In the last few decades, the limits in accelerating gradients of conventional electron accelerators based on metallic cavities prompted considerable efforts in the development of alternative electron acceleration techniques. Hitherto, the acceleration of electrons in the wake of an intense laser pulse propagating in an underdense plasma (Laser Wakefield Acceleration, or LWFA \cite{TajimaDawson79,Malka2002,Esarey2009,Malka2012}) has been proven particularly promising, generating smaller electron accelerators with high accelerating gradients \cite{Mangles2004,Faure2004,Geddes2004}, GeV level final energies \cite{Leemans2014,Gonsalves2019} and femtoseconds duration accelerated beams \cite{Lundh2011}. An important role in this acceleration scheme is played by the technique used to inject relativistic electrons in the accelerating phase of the involved plasma waves \cite{Esarey2009,Malka2012}. Among the numerous demonstrated injection techniques, one of particular simplicity and often chosen is called ionization injection \cite{McGuffey2010,Pak2010,Pollock2011,Chen2012,Golovin2015,Mirzaie2015,Couperus2017,Irman2018}. It consists in using a gas mixture to generate the required plasma. This mixture is mainly composed by a low atomic number Z gas, like hydrogen or helium, that is ionized very early in the laser-gas interaction. The other part of the mixture is composed by a higher Z dopant gas, like nitrogen, whose first levels of ionization are reached early in the laser-gas interaction. However, the higher ionization levels of this gas can be accessed only at higher values of the laser transverse electric field, normally near the peak of the envelope of the laser, i.e. later than the first levels of the high Z gas and the ones of the low Z gas. Tailoring properly the laser and plasma parameters, these ionization levels are reached only during the short period when the laser is near its maximum focusing. It provides a reserve  of electrons that can  be stripped off from the high Z ions near the peak of the  laser pulse ``just in time'' to be trapped in the plasma wave  in the  wake of the laser (this case is referred to as self-truncated ionization injection \cite{Zeng2014,Mirzaie2015,Couperus2017,Irman2018}).

Particle in Cell (PIC) \cite{BirdsallLangdon2004} modeling of this interaction is an essential investigation technique to design and analyze LWFA experiments with ionization injection. The most common technique to take into account tunneling ionization in PIC simulations of these phenomena is to compute at each time step the number of electrons freed from their atoms through the Ammosov–Delone–Krainov direct current (ADK DC) ionization rate \cite{Perelomov1966,ADK1986,Yudin2001}. When a new electron is created in  this way, a sufficiently accurate approximation for momentum conservation is to assign it an initial zero momentum, since the heavy ion from which it is created can be considered immobile. At later times, the electron quickly acquires a transverse momentum whose normalized magnitude is $p_{\perp} (t) = |\mathbf{A}_{\perp} (t)-\mathbf{A}_{\perp} (t_{\rm{ioniz}})|$ \cite{Cowan2011,Gibbon,Macchi} (see Appendix \ref{appendixB}), where $\mathbf{A}_{\perp} (t)$,  $\mathbf{A}_{\perp} (t_{\rm{ioniz}})$ are the instantaneous laser transverse vector potential at time $t$ and at the ionization time $t_{\rm{ioniz}}$. Given the high frequency oscillations of the laser, this results in a quiver motion following the laser oscillations. This occurs through the instantaneous interaction force of the electron with the laser pulse, the Lorentz force, which takes into account the high frequency laser oscillations. In the following, this kind of simulation will be referred to as standard laser simulations, which can be performed in Cartesian geometries or quasi-cylindrical geometry \cite{Lifschitz2009}. This kind of simulations, which needs to be performed in three dimensions (3D) in order to have physical accuracy \cite{Davoine2008}, requires large amounts of resources, due to the disparity between the typical length of a LWFA accelerator, at least of the order of one millimeter, and the smallest scale to resolve, the laser pulse carrier wavelength $\lambda_0$, of the order of one micron. The  parameter space to explore to understand and design LWFA experiments is vast, and cannot be explored directly with numerous standard laser simulations in 3D. For these reasons, reduced models for LWFA simulations are of paramount interest for preliminary studies, because they can significantly reduce the computation time at the cost of introducing physical approximations which are reasonable in many LWFA configurations. One of the most general of these reduced models is the use of azimuthal Fourier decomposition or quasi-cylindrical geometry \cite{Lifschitz2009}, which takes into account only the first azimuthal modes of the electromagnetic fields, reducing the cost of simulation with quasi-3D accuracy to the cost of approximately $N_{m}$ 2D PIC simulations  on a cylindrical grid, where $N_m$ is the number of considered azimuthal modes. For most preliminary studies of LWFA, $N_m=2$ is sufficient, while for the reconstruction of more realistic laser pulses a higher number of modes is necessary \cite{zemzemi2020azimuthal}.

Another reduced model of interest for LWFA is given by the envelope or ponderomotive guiding center model. In most LWFA set-ups, normally the ratio between the scales involved in the excitation of the required plasma waves by the ponderomotive force and the smallest scale to resolve is more than ten. Thus, an averaged formulation of the ponderomotive interaction between the laser pulse and the plasma can be used to obtain accurate results in shorter simulation times, often by one or two orders of magnitude with the same geometry, lifting the requirement of the resolution of the laser wavelength \cite{Mora1997,Gordon2000,Huang2006,Benedetti2010,Cowan2011,Tomassini2017,Terzani2019,Massimo2019,Silva2019}. Besides, preliminary studies of LWFA with this model can be performed in quasi-cylindrical geometry considering only one azimuthal mode representing perfect cylindrical symmetry \cite{Mora1997,Benedetti2010,Tomassini2016MatchingSF,Massimo2019cylindrical}. In that particular geometry, the savings in computational resources are even larger because
a single azimuthal mode is used.

Modeling tunneling ionization in envelope simulations as in standard laser simulations, using the ADK DC ionization rate and the zero-momentum initialization for the new electrons, does not yield accurate results, since the high frequency laser oscillations and thus the electron motion is not well resolved. The residual momentum of the electrons stripped from the atoms/ions after the passage of the laser pulse strongly depends on the extraction field phase, which is normally poorly resolved in an envelope simulation. To circumvent this problem, in the cylindrical envelope code INF\&RNO a reconstruction of the high frequency laser oscillations near the laser pulse is performed at each timestep, calculating the full ionization rate and describing the ionization-quiver dynamics of the new electrons \cite{BenedettiAAC2016}. Although very accurate, a disadvantage of this approach is that an additional grid is  required to interpolate the force  acting on the ions for the ionization and on the  new electrons for their quiver motion. However, if the quiver motion of the electrons does not need to be reconstructed, a reduced approach would ideally reconstruct the main integrated parameters of the electron bunches obtained through ionization injection in LWFA, i.e. charge, average energy, energy spread, rms sizes, normalized emittance. In particular in \cite{Chen2013} it was shown that the use of averaged ADK ionization rate at each timestep of envelope simulations gives a more accurate estimate of the injected charge, and additionally in \cite{Tomassini2017} a procedure to reconstruct the residual transverse momentum spread in envelope simulations, based on the analytical results in \cite{Schroeder2014}, was outlined. The results in \cite{Tomassini2017} have been obtained with values of the maximum normalized vector potential of the laser $a_0=A_{\rm{max}}<1$ . 

For $a_0>1$ the same procedure computes a trapped charge in LWFA envelope PIC simulations that is lower than the one computed in standard laser PIC simulations. Therefore, in this work the reason for this disagreement is discussed, i.e. the initialization of the longitudinal momentum of the electrons created by ionization, and an extension of the tunneling ionization modeling procedure described in \cite{Tomassini2017} is proposed, to obtain accurate results also for $a_0>1$. This extended procedure has been implemented in the open source PIC code \Smilei \cite{Smilei2018,Beck2019}, used for the simulations of the manuscript. In an unified framework, \Smilei can run both standard laser and envelope simulations, in Cartesian geometries \cite{Massimo2019} and quasi-cylindrical geometry \cite{Massimo2019cylindrical}, the latter used for the simulations of this manuscript. The envelope ionization procedure proposed in this manuscript can be used for Cartesian geometries, but also in purely cylindrical geometry (i.e. with only the azimuthal mode $m=0$) and at the same time take into account the initial electron momentum asymmetry intrinsic with a linear polarization for the laser, provided that the wakefields present a significant degree of cylindrical symmetry (i.e. the envelope of the laser has cylindrical symmetry).  Besides, using only one azimuthal mode  in an envelope cylindrical simulation yields a speed-up compared to a standard laser simulation in quasi-cylindrical geometry, where the minimum number of azimuthal modes to use is two \cite{Massimo2019cylindrical}. This is of particular interest for LWFA, where 2D Cartesian simulations fail to give accurate results \cite{Davoine2008} and at least 3D, quasi-cylindrical simulations \cite{Lifschitz2009} or envelope cylindrical simulations are necessary \cite{Massimo2019cylindrical}. The proposed envelope ionization procedure cannot yield an accurate description of LWFA set-ups where carrier-envelope effects play an important role in ionization injection, like in LWFA with few-cycles laser pulses \cite{Lifschitz2012,Faure2019}, where in general the envelope model cannot yield accurate results.

The manuscript is organized as follows: in the second section, the procedure to model ionization in PIC envelope simulations through the ADK model is presented and its characteristic elements are discussed, i.e. the use of the averaged ionization rate, the initialization of the transverse and longitudinal momentum of the newly created electrons. The authors' original contribution is included in the initialization of the longitudinal momentum of the new electrons, a key to obtain accurate results with $a_0>1$. In the third section, a basic ionization benchmark case is introduced to compare the results of a standard PIC simulation and the equivalent envelope simulation, for linear and circular polarization. In the fourth section a well-known 1D model of LWFA is reviewed to hint at the importance of an accurate initialization of the longitudinal momentum of the new electrons like in the proposed ionization procedure. In the fifth section, the comparison between a standard simulation and an envelope simulation with the proposed ionization technique for a full LWFA benchmark with ionization of N$^{5+}$ is presented. Both the benchmarks of the manuscript are run in regimes with $a_0>1$. In the fifth section it is shown how results with sufficient accuracy for preliminary studies can be obtained with the proposed ionization procedure in a very short time. In Appendix \ref{appendixA} the equations of the envelope model used in the manuscript simulations are briefly reviewed. In Appendix \ref{appendixB} the derivation of the initial momentum values assigned to the electrons created with the proposed ionization procedure is described.

\section{Tunneling ionization algorithm with an envelope model}
In the next subsections, the elements of the envelope ionization procedure implemented in \Smilei are outlined, for both laser and envelope simulations. In the following equations of the manuscript, unless specified, normalized units will be used. Charges are normalized to the unit charge $e$, velocities to the speed  of light $c$, masses to the electron mass $m_e$, lengths to the inverse of the laser carrier wavenumber $\lambda_0/2\pi$, momenta by $m_ec$.

\subsection{Tunneling ionization rate with an envelope}\label{ADK_AC_ioniz_rate}
Following the notation in \cite{Nuter2011}, the ADK DC tunneling ionization rate of an atom/ion under the effect of a constant electric field of magnitude $|\mathbf{E}|$  is, in atomic units ($4.134\cdot10^{16}$ Hz) \cite{Perelomov1966,ADK1986,Yudin2001,Nuter2011}:
\begin{equation}\label{ADK_DC}
W_{ADK,\thinspace DC} = A_{n^*,l^*}B_{l,|m|}\thinspace I_p \left( \frac{2(2I_p)^{3/2}}{|\mathbf{E}|} \right)^{2n^*-|m|-1}\rm{exp}\left(-\frac{2}{3}\frac{(2I_p)^{3/2}}{|\mathbf{E}|}\right),
\end{equation}
where the coefficients $A_{n^*,l^*}$, $B_{l,|m|}$ are given by
\begin{equation}
A_{n^*,l^*} = \frac{2^{2n^*}}{n^*\Gamma(n^*+l^*+1)\Gamma(n^*-l^*)},\quad B_{l,|m|} = \frac{(2l+1)(l+|m|)!}{2^{|m|}|m|!(l-|m|)!}.
\end{equation}
In the previous equations, $I_p$ is the ionization potential for the $Z+1$ level of ionization  normalized to $27.2116$ eV, $|\mathbf{E}|$ is normalized to $0.514224$ TV/m, $\Gamma(x)$ is the gamma function, $n^*=Z/\sqrt{2I_p}$, $l^*=n^*-1$ and $l$ and $m$ are the angular momentum and its projection on the laser polarization vector respectively. In \cite{ADK1986} it is shown that the ionization rate for $m=0$ is dominant, thus only this one is taken into account in \Smilei.

In a standard laser simulation with single-level ionization, the probability of ionization at each timestep for each atom/ion is  computed as $1-\rm{exp}(-W_{ADK,\thinspace DC} \Delta t)$, where $\Delta t$ is the integration timestep of the simulation. The extension to multiple-level ionization is described in \cite{Nuter2011}.
The use of the ADK DC rate in a standard PIC is physically justified from the fact that normally the field does not vary significantly within an interval $\Delta t$, chosen to resolve the laser oscillation frequency. As discussed in \cite{Chen2013}, this approximation is no longer valid in an envelope simulation, where the integration timestep can potentially contain multiple laser oscillations. Thus, as recommended in the same reference, the averaged version of the ionization rate in alternate current $W_{ADK,\thinspace AC}$ must be used with an envelope model for accurate results. For circular polarization, since the laser electric field magnitude does not change within a laser oscillation, $W_{ADK,\thinspace AC} = W_{ADK,\thinspace DC}$. For linear polarization, the averaging of $W_{ADK,\thinspace DC}$ yields \cite{ADK1986,Chen2013}:
\begin{equation}\label{ADK_AC}
W_{ADK,\thinspace AC} = \left[ \frac{3}{\pi}  |E_{\rm{envelope}}| \thinspace  (2I_p)^{-3/2}\right]^{1/2} \thinspace W_{ADK,\thinspace DC} (|E_{\rm{envelope}}|).
\end{equation}

For the computation of the AC ionization rate, the magnitude of the laser envelope electric field $|E_{\rm{envelope}}|$, including the longitudinal and transverse field components (see Appendix \ref{appendixA} for their computation) must be used  instead  of the instantaneous field $|\mathbf{E}|$ used in  Eq. \ref{ADK_DC}. Note that the ponderomotive force does not change the ionization rate \cite{Delone98}.

The use of  $W_{ADK,\thinspace AC}$ in envelope simulations ensures more correct computations of the total amount of electrons created by tunneling ionization \cite{ADK1986,Tomassini2017}. However, as discussed in the next subsections, to correctly model LWFA, also a correct initialization of the transverse and longitudinal momenta of these electrons is needed with high laser intensities.

\subsection{Transverse momentum initialization}\label{p_perp_initialization}
In standard laser simulations, to ensure the conservation of momentum, the momentum assigned to the new eletrons created by ionization is normally the same of the atom/ion from which they originated. In most LWFA set-ups, this momentum is initially zero. The new electrons almost instantly acquire a quickly oscillating component in the transverse momentum in their interaction with the laser (see Appendix \ref{appendixB}), which in general can be non-zero and depends on the position/phase of the electron. Thus, these electrons initially have a certain spread in the transverse momentum, quantified in \cite{Schroeder2014}. Neglecting the high frequency oscillations, their averaged dynamics is then determined by the ponderomotive force of the laser and in case also by the averaged plasma wakefield in LWFA.

The aim of the proposed ionization procedure for an envelope simulation is not necessarily to obtain injected electron bunches identical in the phase space to those in an equivalent standard laser simulation, but to obtain residual statistical properties that are at least similar. Here the term residual denotes the values of the physical quantities computed through the averaging over the laser oscillations, but taking into account their initial momentum spread.
To satisfy this requirement, in \Smilei the same procedure for transverse momentum initialization of the new electrons in envelope simulations as in the PIC code ALaDyn \cite{ALaDyn2008,Terzani2019} and the hybrid PIC-fluid quasi-static code QFluid \cite{Tomassini2017} was implemented, for both linear and circular polarizations. This procedure is briefly reviewed in the following and discussed in detail in Appendix \ref{appendixB}.

To preliminary define the notation used in this work (the same notation from \cite{Cowan2011}, used also in the Appendices), the transverse vector potential of a laser that can be described by a laser envelope is denoted with $\mathbf{\hat{A}_\perp(\mathbf{x},t)}=\Re\left[\mathbf{\tilde{A}}_\perp(\mathbf{x},t) e^{i(x-t)}\right]$, where $\mathbf{\tilde{A}}_\perp$ is the complex envelope of the laser transverse vector potential. This envelope, whose evolution is described by the envelope equation (see Appendix \ref{appendixA}) takes the form $\mathbf{\tilde{A}}_\perp = \mathbf{e}_y\tilde{A}$ for a laser linearly polarized in the $y$ direction and $\mathbf{\tilde{A}}_\perp = \frac{\tilde{A}}{\sqrt{2}}(\mathbf{e}_y\pm i\mathbf{e}_z)$ for a circularly polarized laser. In \Smilei, at a given timestep the envelope equation is solved for $\tilde{A}$ for linear polarization and for $\tilde{A}/\sqrt{2}$ for circular polarization (see Appendix \ref{appendixA}).

Electrons stripped from an atom/ion by a laser with linear polarization have an initial Gaussian distribution in the transverse momentum $p_{pol}$ in the polarization direction, with rms width \cite{Schroeder2014} 
\begin{equation}\label{transverse_momentum_initialization}
\sigma_{p_{pol}} = |\tilde{A}|\cdot \left(\frac{3}{2}|\tilde{E}_{\rm{envelope}}|\right)^{1/2}(2I_p)^{-3/4},
\end{equation}
where $|\tilde{A}|$ is the magnitude of the complex envelope of the laser vector potential in the polarization direction. Thus, for linear polarization simulations, the initial transverse momentum in the polarization direction of the new electron $p_{pol,0}$ is assigned as a pseudorandom number drawn from a Gaussian distribution of rms width $\sigma_{p_{pol}}$, computed from Eq. \ref{transverse_momentum_initialization} using the $|\tilde{A}|$ interpolated at the new electron position (i.e. the same position of the atom/ion from which it originated). Since the laser is propagating in the $x$ direction, the initial momentum in the direction perpendicular to $x$ and to the polarization direction is assigned as zero. Thus, in linear polarization the magnitude of the initial transverse momentum of the new electrons is $|\mathbf{p}_{\perp,0}|=p_{pol,0}$.
We remark that, using the notation in \cite{Schroeder2014}, Eq. \ref{transverse_momentum_initialization} is valid when 
\begin{equation}\label{eqref:Delta}
\Delta = \left(\frac{3}{2}|\tilde{E}_{\rm{envelope}}|\right)^{1/2}(2I_p)^{-3/4} \ll 1.
\end{equation}
For example, the normalized $I_p$ corresponding to the ionization of the sixth [seventh] level of nitrogen (used in the benchmarks of this work) are equal to $552\thinspace \rm{eV}/13.6 \thinspace \rm{eV}=40.6$ [$667\thinspace \rm{eV} /13.6\thinspace \rm{eV}=49$], thus $\Delta \ll 1$ as long as $|\tilde{E}_{\rm{envelope}}|\ll (2/3)(2I_p)^{3/2} = 28\thinspace [34]$ TV/m, corresponding to values of the normalized maximum vector potential $a_0\ll 7\thinspace [8.5]$. These conditions are satisfied in the benchmarks of this work.\\
For circular polarization, since the magnitude of the laser transverse vector potential does not change within a timestep, the transverse momentum of the new electron is assigned as $|\mathbf{p}_{\perp,0}|=|\mathbf{\hat{p}}_\perp|=|\mathbf{\hat{A}}_\perp|=|\tilde{A}|/\sqrt{2}$, with an azimuthal angle randomly extracted from $0$ to $2\pi$. Again $|\tilde{A}|$ is interpolated at the new electron position at the iteration when ionization occurs.

\subsection{Longitudinal momentum initialization}\label{px_initialization}
In standard laser simulations, also the longitudinal momentum assigned to the new eletrons created by ionization is typically the same as the atom/ion which they originated from, in general zero for LWFA set-ups. As in the transverse case, the new electrons almost instantly acquire a quickly oscillating component in the longitudinal momentum in their interaction with the laser (see Appendix \ref{appendixB}), which depends on the vector potential to which the electron is subject. Considering that also the transverse momentum depends on the vector potential, the two momentum components are not independent. Electrons subject to a higher $|\mathbf{\hat{A}}_\perp|$ should have both higher transverse and longitudinal momentum at the same time. Besides, since the residual transverse momentum spread is not zero, the initial longitudinal momentum spread will be in general non zero. To the authors' knowledge, in literature there is no analytical result that estimates this spread. Additionally, the calculation of the average residual $p_x$ implies the averaging over the square of a locally sinusoidal function (see Appendix \ref{appendixB}) hence it can be  inferred that in an envelope simulation a residual positive average longitudinal momentum (a drift in the positive $x$ direction) should be present. 

For values of $a_0<1$, the use of the ADK AC ionization rate and the initialization of the transverse momentum of the new electrons described in the previous section are sufficient to have a statistically accurate description of the electrons obtained from ionization in an envelope ionization \cite{Tomassini2017}. The ionization process $Ar^{8+} \rightarrow Ar^{9+}$ requires $a_0=0.4$ for $\lambda_0=0.4$ $\mu$m for example, and can be correctly modeled with this approach \cite{Tomassini2017}. However, to ionize some dopants commonly used in LWFA with ionization injection, higher values of $a_0$ are necessary. In the next sections it is shown that in these cases also an initialization of the longitudinal momentum of the new electrons is necessary. For example, the commonly chosen process N$^{5+} \rightarrow $N$^{6+}$ requires $a_0>1.8$ for $\lambda_0=0.8$ $\mu$m. This process is used in order to benchmark the momentum initialization that is proposed in this work, as shown in the next sections.

The initial $p_{x,0}$ of the new electrons can be chosen to be initialized as
\begin{equation}
p_{x,0}= \left\{
\begin{array}{ll}
      |\tilde{A}|^2/4+p_{pol,0}^2/2 & \text{for linear polarization} \\
      |\tilde{A}|^2/2 & \text{for circular polarization}. \\
\end{array}
\right. 
\end{equation}
This choice, motivated in Appendix \ref{appendixB}, yields statistical characteristics in the longitudinal momentum that are similar to those of the electrons created through tunneling ionization within a timestep $\Delta t$, as it is shown in the following sections. A worthwile remark is that the maximum values assigned to $p_{x,0}$ scale as $a_0^2/2$ or smaller, thus they start to become relevant only for values of $a_0$ approaching 1. In the following sections it is shown also that for $a_0>1$ this choice has effects also on the transverse momentum evolution, due to the contribution of $p_x$ to the relativistic inertia of the electrons, quantified by their Lorentz factor. In LWFA, this choice of $p_{x,0}$ also allows the new electrons to have an averaged momentum in the $x$ direction high enough to be trapped and efficently accelerated by the plasma wave behind the laser (see sections \ref{1Dmodel}, \ref{LWFAbenchmark}).

\section{Basic case study: Ionization of nitrogen slab}\label{nitrogen_benchmark}
LWFA with ionization injection with high $a_0$ is rich of nonlinear phenomena, which take place contemporarily with the further ionization of the high Z gas: relativistic plasma wave excitation, relativistic self-focusing \cite{Sun1987}, trapping  and acceleration of electrons. To show more in detail the comparison in the momentum space obtained between the two kinds of simulation, standard laser PIC simulation and envelope simulation, a benchmark is presented in this section, where the interaction of the new electrons with the laser is limited, no plasma wave is present to accelerate electrons and the laser pulse evolution is negligible compared to that in vacuum.

\begin{figure}[htbp]
\begin{center}
\includegraphics[scale=0.55]{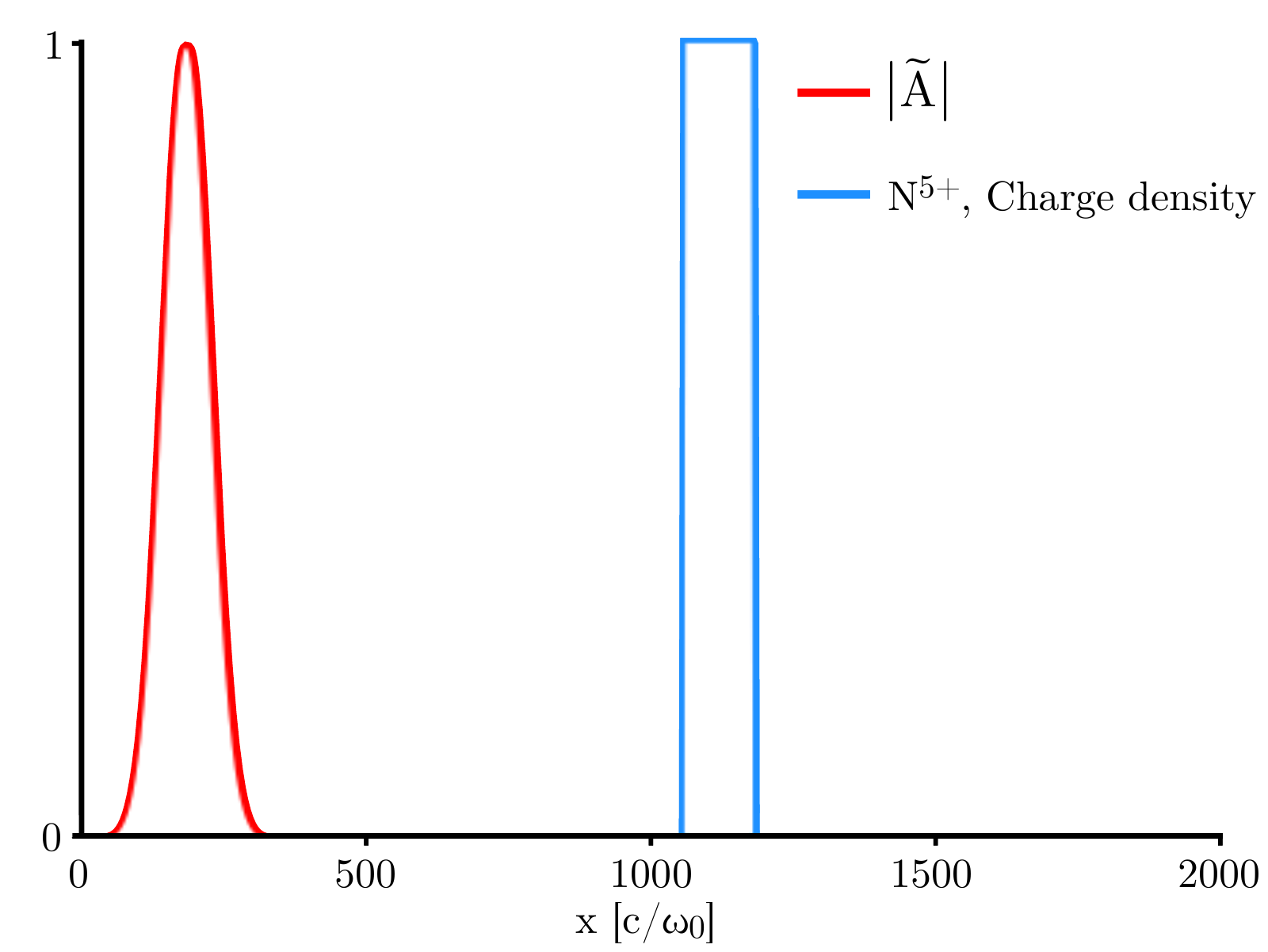}
\caption{Initial state of the  N$^{5+}$ ionization benchmark in the envelope simulations along the propagation axis $x$. The envelope module $|\tilde{A}|$ and the charge density of the N$^{5+}$ slab are normalized to 1.}
\label{fig:nitrogen_slab_initial_state}
\end{center}
\end{figure}

This simple benchmark consists in initializing an immobile cylindrical slab of N$^{5+}$ ions and their neutralizing electrons (those obtained ionizing the first 5 atomic levels) in vacuum and let the ions be further ionized by the passage of an intense Gaussian laser pulse with $a_0=2$ in the linear polarization simulation and with $a_0=2\sqrt{2}$ in the circular polarization simulation (see Fig. \ref{fig:nitrogen_slab_initial_state}). The nitrogen slab is a cylinder of radius $R=100$ $\lambda_0/2\pi$ and length $L=120$ $\lambda_0/2\pi$, whose axis coincides with the laser propagation axis (the $x$ axis). The longitudinal density profile of the nitrogen slab is a plateau with density $n_0=0.0000056$ $n_c$, where $n_c$ is the critical density for the chosen laser wavelength $\lambda_0$, with an upramp and a downramp of zero length.

The Gaussian laser pulse, propagating along the positive $x$ direction with carrier wavelength $\lambda_0=0.8$ $\mu$m (as for a Ti:Sa laser system), has a waist size $w_0=100$  $\lambda_0/2\pi$ and full width half maximum duration in field $L_{FWHM}=92.5$  $\lambda_0/2\pi c$. The laser pulse initial position at the beginning of the simulation is at a distance $875$ $\lambda_0/2\pi$ from the nitrogen slab along the $x$ axis and its focal plane is placed at the beginning of the nitrogen slab. In the linear polarization simulations, the polarization direction of the laser is the $y$ direction.

The laser pulse passes through the N$^{5+}$ slab triggering injection and the new electrons are left to move subject to the initial ponderomotive force of the laser and to the electric field that is created from the charge imbalance progressively created in the slab after the ionization and the movement of the electrons. 

After a time $T = 1277.2$ $\lambda_0/2\pi c$, when the laser is far from the N$^{5+}$ slab, the momentum distributions of the newly created electrons in the envelope and standard laser simulations are compared. It is important to remark that this comparison can show an agreement only after the interaction with the laser is finished. Indeed, the transverse momentum distribution in the standard laser simulation when the laser is still in the nitrogen slab would show two peaks due to the quivering motion, which would be of course completely absent in the envelope simulation. In other words, the momenta in the envelope simulations represent the slow-varying part $\mathbf{\bar{p}}$ of the real momenta $\mathbf{p}=\mathbf{\bar{p}}+\mathbf{\hat{p}}$ (see Appendix \ref{appendixA}). After the interaction with the laser, the quickly oscillating part of the momenta $\mathbf{\hat{p}}$ is negligible and the momenta of the envelope simulations $\mathbf{\bar{p}}$ can be compared to the total momenta of the standard laser simulation $\mathbf{p}$. For this reason, the bar over the momenta in the envelope simulations results is dropped in the following for the sake of brevity.

Both the simulation types, standard laser and envelope, are run in quasi-cylindrical geometry  \cite{Lifschitz2009}, with transverse resolution $\Delta r =2$ $\lambda_0/2\pi$. The standard laser simulations model the laser-plasma interaction using two azimuthal modes (modes $m=0,\thinspace 1$), while the envelope simulations does it with only one azimuthal mode ($m=0$, representing perfect cylindrical symmetry). Although in the simulation with linear polarization the cylindrical symmetry is not respected in the momentum initialization of the envelope simulation, the resulting asymmetries do not seem to lead to significant statistical differences in the momentum distribution of the electrons, since the magnitude of the currents and field with $m=1$  is negligible compared to those with $m=0$. The standard laser simulations have a longitudinal resolution $\Delta x_{\rm{laser}}=0.125$ $\lambda_0/2\pi$ and integration timestep $\Delta t_{\rm{laser}}=0.124$  $\lambda_0/2\pi c$, while in the envelope simulations these ones are $\Delta x_{\rm{envelope}}=0.8$ $\lambda_0/2\pi$ and $\Delta t_{\rm{envelope}}=0.79825\Delta x_{\rm{envelope}}/c$. With this choice of integration timestep, the envelope solver is still stable \cite{Massimo2019} and $\Delta t_{\rm{envelope}}=(515/10)$, thus the results of the two kinds of simulation can be easily compared at the same time, provided that the ratio between the iterations is the same for the standard laser and  envelope simulations. 

The laser envelope is initialized already in the simulation domain, while the laser enters from the left border of the  simulation domain in the  standard laser simulation, through  Silver-M\"uller boundary conditions \cite{He1999,Barucq1997}. Despite this difference, it was ensured that the laser focal plane position and the initial distance between the laser temporal center and the N$^{5+}$ slab was the same for envelope and standard laser simulation. 

The total number of macro-particles per cell is the same in all simulations, $N_x\times N_r\times N_{\theta}=32$, placed regularly along the three directions. In the laser simulation, the distribution of  the macro-particles is $[N_x,\thinspace N_r,\thinspace N_\theta]=[1,\thinspace 4,\thinspace8]$, where $N_x$ and $N_r$ are the number of particles in the $x$ and $r$ directions and $N_\theta$  is the number of particles evenly distributed in the azimuthal angle interval between $0$ and $2\pi$. For the envelope simulations, the macro-particles distribution is $[N_x,\thinspace N_r,\thinspace N_\theta]=[8,\thinspace 4,\thinspace1]$.  Since only one azimuthal mode is used in the envelope simulations, i.e. the physical phenomena are assumed to be cylindrically symmetric, in the envelope simulations $N_\theta=1$ can be used.

The N$^{5+}$ ions are immobile, i.e. although they project charge on the grid and the electromagnetic field and the laser envelope are interpolated at their positions for the ionization operations, their positions and momenta are not changed. This is a reasonable approximation in most LWFA regimes, since the timescales of ion motion are much longer than those of electron motion, and avoids adding a significant computation time for operations which do not impact the final results.

\begin{figure}[htbp]
\begin{center}
\includegraphics[scale=0.15]{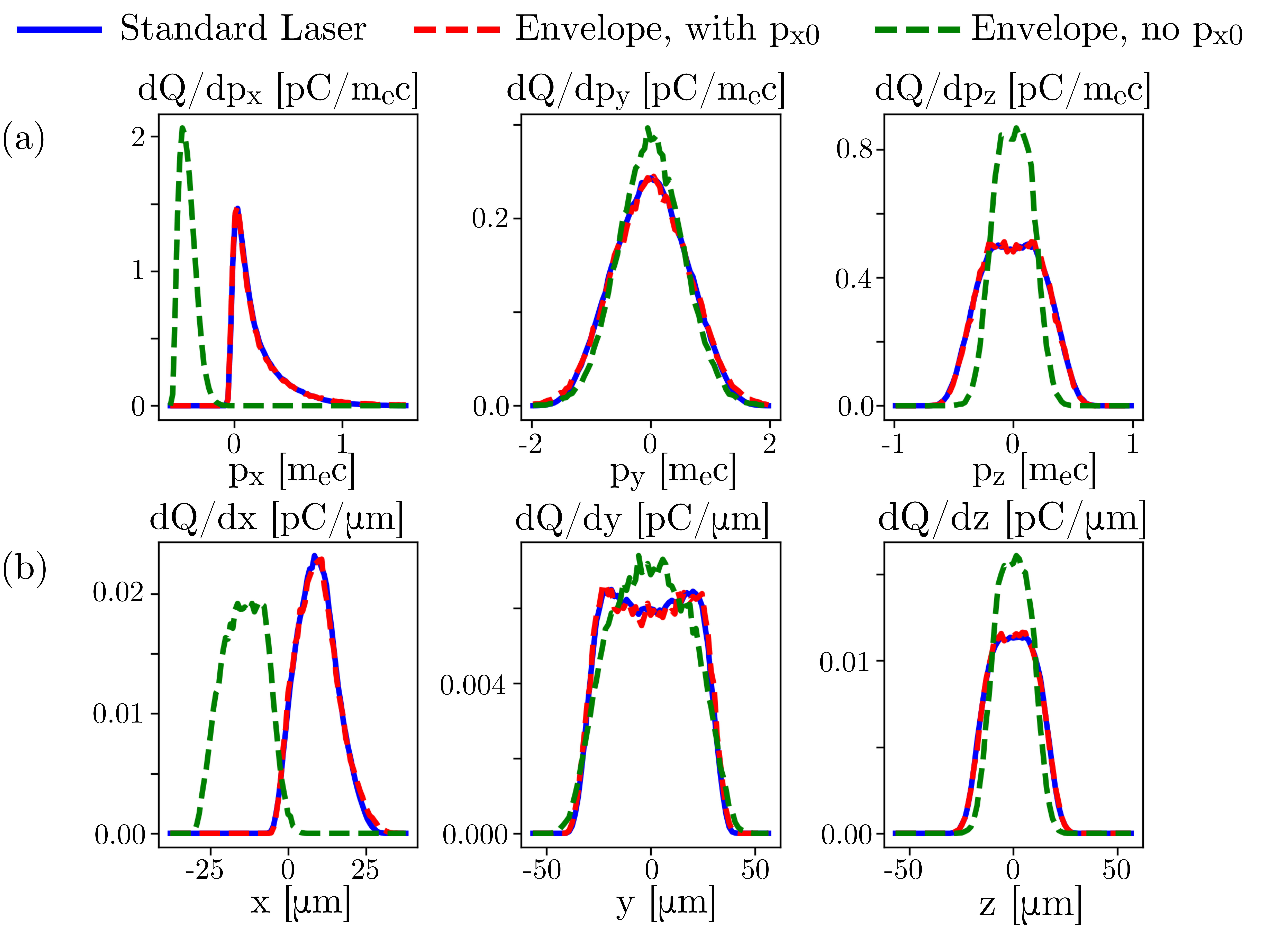}
\caption{\textcolor{black}{Comparison of (a) the momentum distributions and (b) the position distributions of the electrons created by ionization in the N$^{5+}$ benchmark at time $T = 1277.2$ $\lambda_0/2\pi c$, with linear polarization along the $y$ axis.}}
\label{fig:nitrogen_slab_linear_polarization}
\end{center}
\end{figure}

\begin{figure}[htbp]
\begin{center}
\includegraphics[scale=0.15]{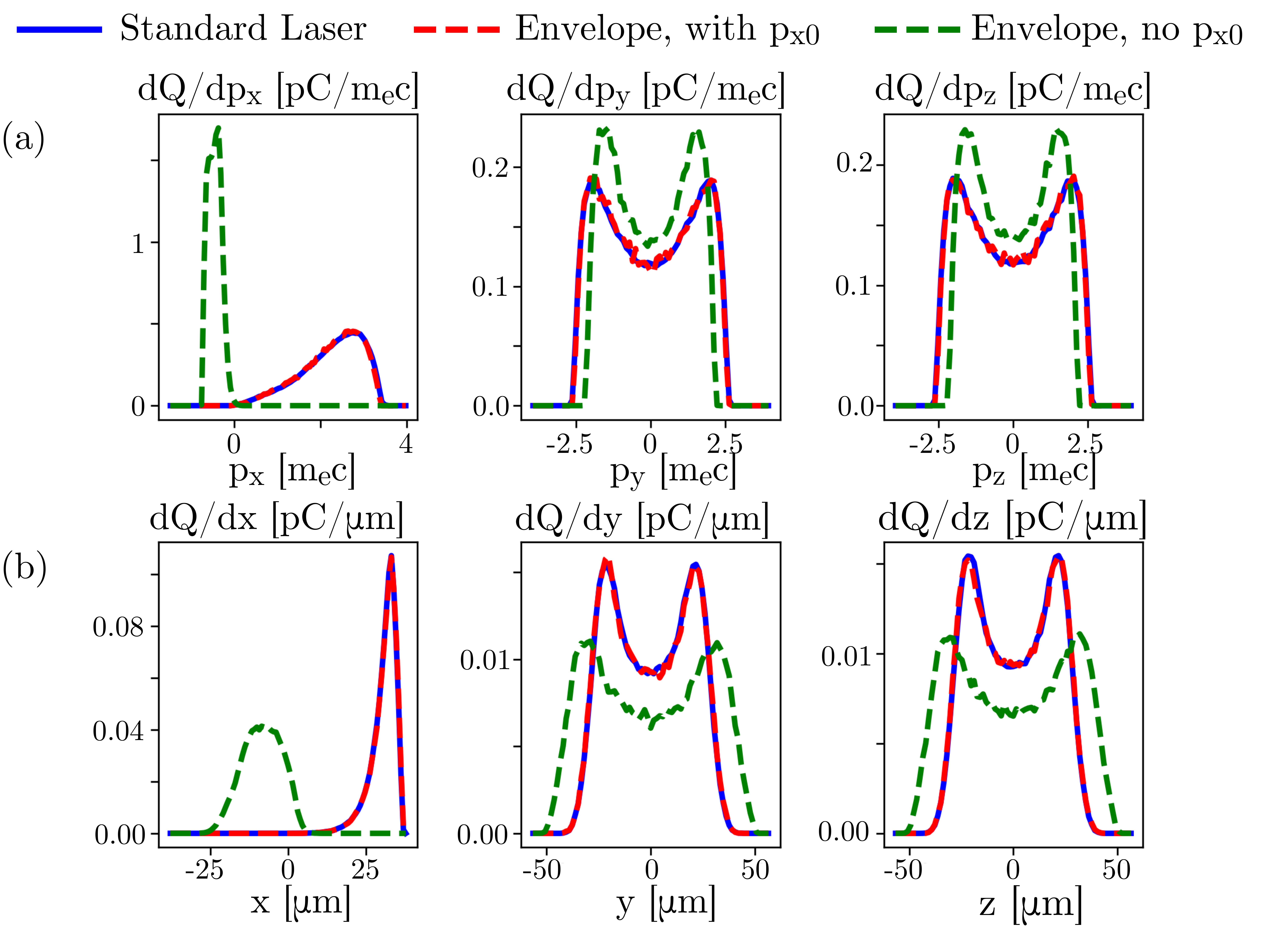}
\caption{\textcolor{black}{Comparison of (a) the momentum distributions and (b) the position distributions of the electrons created by ionization in the N$^{5+}$ benchmark at time $T = 1277.2$ $\lambda_0/2\pi c$, with circular polatization.}}
\label{fig:nitrogen_slab_circular_polarization}
\end{center}
\end{figure}

To show the effects of the $p_x$ initialization for the new electrons in the envelope simulations presented in section \ref{px_initialization}, in the top panels of Figs. \ref{fig:nitrogen_slab_linear_polarization}, \ref{fig:nitrogen_slab_circular_polarization} the comparison of the electron momentum distributions at time $T$ (i.e. 2000 iterations for the envelope simulations, 10300 iterations for the standard laser simulations) is reported. The Figures also report the results obtained without longitudinal momentum initialization, with linear polarization along the $y$ axis and circular polarization. Without this initialization, the longitudinal momentum distribution results are completely different in the two simulations and the transverse momenta distributions display differences as well, although the shape of these distributions is the same as with longitudinal momentum distribution. The electrons in the envelope simulation without $p_x$ initialization seem to be much slower than in the standard laser simulations, pushed on the left by the field created by the charge separation that progressively forms and the laser ponderomotive force. For LWFA, this difference is particularly critical because if the electrons obtained from the dopant do not possess an averaged longitudinal momentum high enough to be trapped in the plasma wave in the wake of the laser, the injected charge will be significantly lower in the envelope simulation (see sections \ref{1Dmodel}, \ref{LWFAbenchmark}). Thus, as it will be shown in the next sections, an agreement can be expected in the injected charge with the proposed $p_x$ initialization procedure in a LWFA simulation with $a_0>1$. Although the transverse momenta initialization is the same for the two envelope simulations (the one described in section \ref{p_perp_initialization}), the resulting distribution of the transverse momenta at time T is quantitatively different, because with the $p_x$ initialization a portion of macro-particles starts to become nearly relativistic ($1+|\mathbf{p}|^2=1+p_x^2+|\mathbf{p}_\perp|^2|\gtrsim 2$), changing their inertia towards the ponderomotive force and the force created by the charge separation. \\

Even in this simple benchmark the difference in momenta at the moment of creation of the new electrons in the envelope simulations results in completely different position distributions at later times. In the bottom panel of Figs. \ref{fig:nitrogen_slab_linear_polarization}, \ref{fig:nitrogen_slab_circular_polarization} the comparison of the electron position distribution at time $T$ is reported. The positions are referred to the center of the N$^{5+}$ slab. As it was evident from the top panels, in the envelope simulaton without $p_x$ initialization the drift in the $x$ direction is lower, thus the macro-particles on average are moving towards the negative $x$ direction, while in the standard laser simulation and in the envelope simulation with $p_x$ initialization they are moving in the opposite direction. The transverse position distributions are different as well.

\section{Electron trapping from LWFA with ionization injection}\label{1Dmodel}
\begin{figure}[htbp]
\begin{center}
\includegraphics[scale=0.8]{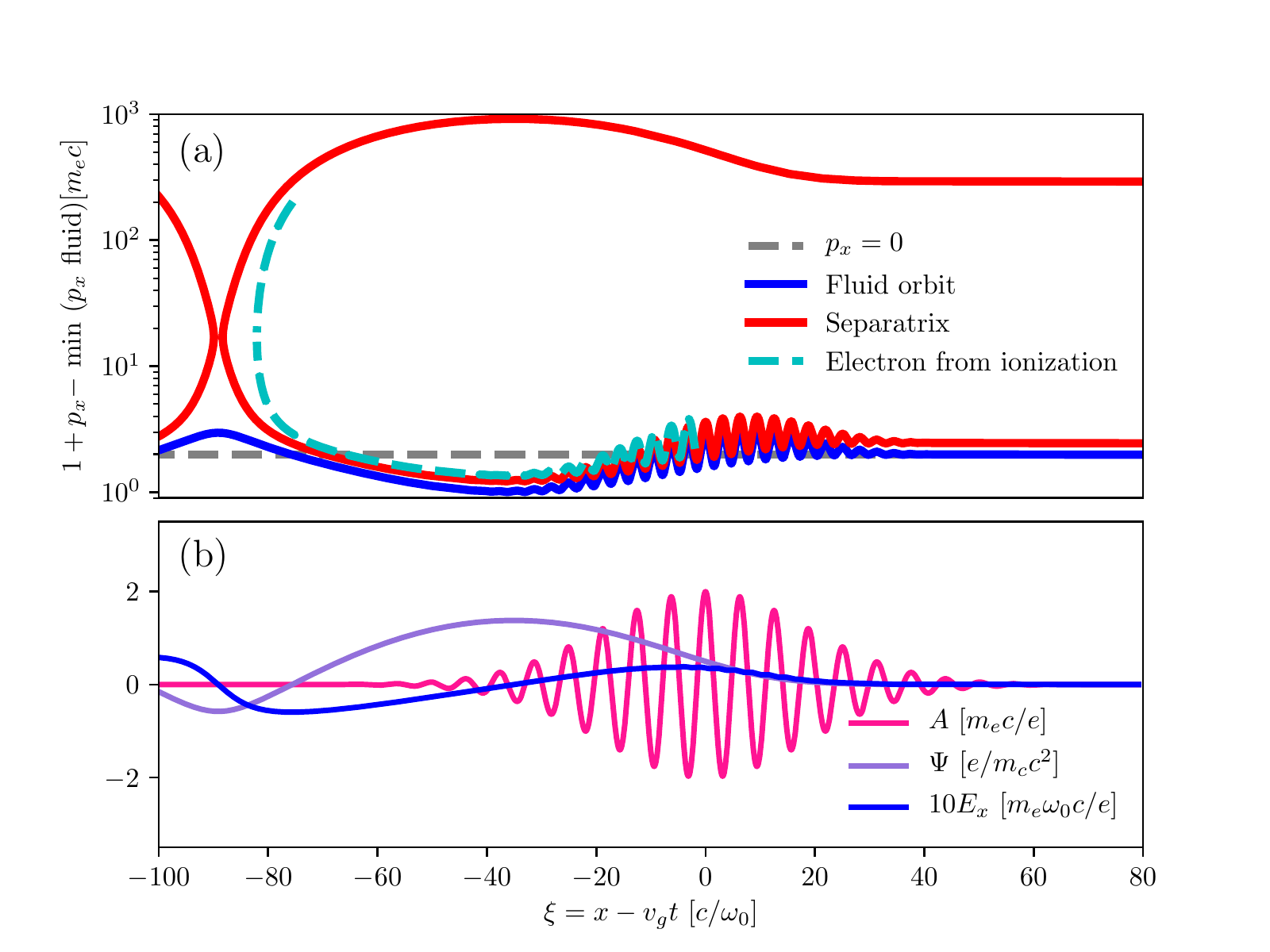}
\caption{Adapted from Figs. 2, 5 of of \cite{FaureCAS} with the almost the same laser and plasma parameters: $a_0=2$, $n_0=0.44\%\thinspace n_c$, $L_{\rm{FWHM}}=44$ $\lambda_0/2\pi$.
\textcolor{black}{(a)} Phase space trajectories of test electrons, including a fluid orbit, the separatrix and the trajectory of an electron stripped from its atom/ion through ionization; \textcolor{black}{(b)} laser vector potential $A$, electrostatic potential $\Psi$ and longitudinal electric field $E_x$. }
\label{fig:phase_space_orbits}
\end{center}
\end{figure}

Before showing the results of an envelope LWFA simulation, a brief review of a well-known 1D model of the dynamics of an electron in a plasma wave \cite{Esarey1995,Esarey2009,Chen2012,FaureCAS} can hint at the importance of an accurate $p_x$ initialization. 
In this model a plasma of constant density $n_0$ is modeled as a cold relativistic fluid with immobile ions. The driver of the plasma oscillations is a linearly polarized laser pulse with carrier wavelength $\lambda_0$, in the quasi-static approximation \cite{Mora1997} described by a transverse vector potential $A(\xi=x-v_gt)=a_0\exp(-\xi^2/\sigma)\cos(\xi)$, where $\sigma=0.5(L_{\rm{FWHM}})^2/\ln2$ ($L_{\rm{FWHM}}$ is the laser pulse FWHM duration in field). \textcolor{black}{The plasma wave phase velocity is assumed to coincide with the driver laser pulse group velocity in the plasma $v_g$.} The laser pulse excites a wakefield with electrostatic potential $\Psi$ and longitudinal electric field $E_x=-\partial_\xi\Psi$ described by the differential equation  \cite{Berezhiani_1992,Esarey1993,Teychenne1993}
\begin{equation}\label{potential_equation}
\partial_\xi^2\Psi=\left[\beta_p\left(1-\frac{(1+A^2)}{\gamma_p^2(1+\Psi)^2}\right)^{-1/2}-1\right],
\end{equation}
where $\gamma_p=(1-\beta_p^2)^{-1/2}=(n_0)^{-1/2}$ is the Lorentz factor associated to the plasma wave phase/group normalized velocity $\beta_p$. Once Eq. \ref{potential_equation} is solved assuming $\Psi\vert_{\xi\rightarrow+\infty}=\partial_\xi\Psi\vert_{\xi\rightarrow+\infty}=0$, through a second-type generating function a conserved Hamiltonian function $H$ can be defined to describe the dynamics of a test electron in this system \cite{Esarey1995,FaureCAS}:
\textcolor{black}{
\begin{equation}\label{hamiltonian}
H=\sqrt{1+p_x^2+p_{\perp}^2}-\Psi-\beta_p p_x=\sqrt{1+p_x^2+|A-A(\xi_0)|^2}-\Psi-\beta_p p_x.
\end{equation}
The last identity was obtained using conservation of transverse canonical momentum $p_\perp-A=p_\perp(\xi_0)-A(\xi_0)$ (since the Hamiltonian is not explicitly dependent on the transverse coordinate $x_\perp$, see also Appendix \ref{appendixB}) and $p_\perp(\xi_0)=0$. For an electron created at rest through ionization, $\xi_0$ is the point where it is released from its atom/ion and $A(\xi_0)=A_{ioniz}\rightarrow p_{\perp}=A-A_{ioniz}$. Free background electrons at rest far from the laser pulse, at $\xi_0\rightarrow+\infty$, have an initial condition $A(\xi_0)=0\rightarrow p_\perp=A$.}
The curves of constant $H$ in the phase space $\xi-p_x$ describe the evolution of the test electron momentum, and give insightful information on its trapping state. Given a value $H_0$ of the Hamiltonian, the evolution of the test electron momentum for each value of $\xi$ can be found inverting Eq. \ref{hamiltonian}. \textcolor{black}{For a  background electron, i.e. $A(\xi_0\rightarrow+\infty)=0\rightarrow p_\perp=A$:}
\begin{equation}
p_x=\beta_p\gamma_p^2(H_0+\Psi)\pm\gamma_p\sqrt{\gamma_p^2(H_0+\Psi)^2-(1+A^2)^2}.
\end{equation}
Electrons with $p_x\vert_{\xi\rightarrow+\infty}=0$  are associated to fluid orbits with $H_0=H_{\rm{fluid}}=1$, describing an electron indefinetely drifting in the negative $\xi$ direction and  oscillating with the plasma wave (blue line of Fig. \ref{fig:phase_space_orbits}). In this sense, the fluid orbit is an untrapped orbit. A trajectory which remains in the plasma wave bucket behind the laser pulse is referred to as a trapped orbit, which characterizes electrons that can be accelerated by the plasma wave. In this system a separatrix curve (red line of Fig. \ref{fig:phase_space_orbits}) with $H_{\rm{sep}}=\sqrt{1+\min(A^2)}/\gamma_p-\min(\Psi)$ separates the untrapped ($H_0>H_{\rm{sep}}$) and trapped orbits ($H_0<H_{\rm{sep}}$). 

This simple model can give some insight on the ionization injection process, as discussed thoroughly in \cite{Chen2012,FaureCAS}. If a test electron is considered as initially at rest and stripped through ionization from a high Z gas atom/ion within the laser pulse at position $\xi_0$, in particular near a peak of the electric field (where $A(\xi_0)\approx0\rightarrow p_\perp=A$ since $p_\perp-A=\text{const}$), the same Hamiltonian of Eq. \ref{hamiltonian} can be used to describe its motion, with value $H_{e}=1-\Psi(\xi_0)$. The ionization injection scheme of LWFA relies on creating through ionization enough electrons in trapped orbits ($H_e<H_{\rm{sep}}$) to accelerate them. Figure \ref{fig:phase_space_orbits} shows the trajectory (cyan line) of such a test electron created through ionization ($\xi_0\approx0$) in a trapped orbit of a nonlinear wakefield ($a_0=2$). Although its initial longitudinal momentum is zero, as discussed in the previous section it acquires a momentum under the effect of the laser and it is trapped in the plasma wave to be then accelerated. In a standard laser simulation of this phenomenon it is accurate to set the $p_{x,0}$ of these electrons as zero, since most of them are created near the peaks  of the electric field, where the vector potential is near zero \cite{FaureCAS}. However, although the electron $p_x$ starts oscillating, the average value of these oscillations is non-zero. Thus, in an envelope simulation where the real laser field varies significantly in an integration timestep it is necessary to initialize a non-zero averaged $p_{x,0}$ to be physically accurate. An averaged $p_{x,0}=0$ would make the electron move outside the separatrix curve, preventing its trapping. 

Additionally, as it was shown in the  previous section, in a multi-dimensional system the correct initialization of $p_{x,0}$ with $a_0>1$ ensures an accurate evaluation of the electron averaged Lorentz factor after its interaction with the laser, which determines its inertia towards the laser ponderomotive force, the forces present in the plasma and the correct description of the transverse coordinates evolution.

\section{Benchmark case study: Laser Wakefield Acceleration with ionization injection}\label{LWFAbenchmark}
In this section, a full LWFA simulation with ionization injection is presented as benchmark. A Gaussian laser pulse of wavelength $\lambda_0=0.8$ $\mu$m, propagating in the positive $x$ direction and linearly polarized in the $y$ direction with $a_0=2.5$, is focused at the beginning of a target made by an already ionized mixture of  $99\%$ He and $1\%$ N$^{5+}$, exciting a wakefield and further ionizing the nitrogen. Some electrons extracted from the last two ionization levels of nitrogen are then trapped and accelerated in the laser wakefield. The results of two envelope simulations are shown, simulations whose only difference is that one of them uses the $p_x$ initialization described in section \ref{px_initialization}. 
They are benchmarked against a standard laser simulation.
The ADK AC ionization rate and the initialization of the transverse momentum of the electrons created by ionization in the envelope simulations are those reported in sections \ref{ADK_AC_ioniz_rate}, \ref{p_perp_initialization}. As in the simulations of section \ref{nitrogen_benchmark}, the ions are immobile in all simulations.

The laser pulse has a waist $w_0 = 18.7$ $\mu$m and a full width half maximum duration in intensity $L_{FWHM} = 33$ fs. The N$^{5+}$ target longitudinal profile density is $n_0=3.4\cdot10^{18}$ cm$^{-3}$, with a linear upramp of length $100$ $\lambda_0/2\pi$. The target has a radius  $R=360$ $\lambda_0/2\pi$ and constant radial density profile. The laser focal plane is placed at the beginning of the target.

For the reader's convenience, the numerical parameters of the laser and envelope simulations are reported in Table \ref{table:sim_params_benchmark_LWFA}. As in the previous section benchmark, all the simulations have the same transverse resolution. The moving window physical size is the same, but $\Delta x_{\rm{envelope}}=8\thinspace \Delta x_{\rm{laser}}$ and $\Delta t_{\rm{envelope}}=0.82749 \thinspace\Delta x_{\rm{envelope}}/c$ respectively. The timestep choice ensures that $\Delta t_{\rm{envelope}}/\Delta t_{\rm{laser}}\approx 20/3$, allowing to easily compare the results at approximately the same time, provided that the number of iterations of the two kind of simulations has the same ratio. The same macro-particle distribution used for the benchmark in section \ref{nitrogen_benchmark} was chosen. Since $\Delta x_{\rm{envelope}}=8\thinspace \Delta x_{\rm{laser}}$, the standard laser and envelope simulations have the same spatial sampling $N_x\Delta x\times N_r\Delta r$ in the $x-r$ plane.

\begin{table}[hbtp]
\centering
\renewcommand{\arraystretch}{1}
\begin{tabular}{l|c|c|}
\cline{2-3}
&  \multicolumn{1}{l|}{Standard Laser Simulation} & \multicolumn{1}{l|}{Envelope Simulations} \\ \hline
\multicolumn{1}{|l|}{$\Delta x$ [$\lambda_0/2\pi$]}  & 0.125 & 1 \\ \hline
\multicolumn{1}{|l|}{$c\Delta t/\Delta x$ }  & 0.993 & 0.82749 \\ \hline
\multicolumn{1}{|l|}{N$_{\rm{modes}}$}  & 2 & 1 \\ \hline
\multicolumn{1}{|l|}{$N_{\rm{cells},x}$}  & 3328 & 416 \\ \hline
\multicolumn{1}{|l|}{$N_{\text{macro-particles per cell}}[N_x,\thinspace N_r,\thinspace N_\theta]$}  & 1,4,8 & 8,4,1 \\ \hline
\end{tabular}
\begin{tabular}{l|c|c|}
\cline{2-2}
&  \multicolumn{1}{l|}{All Simulations} \\ \hline
\multicolumn{1}{|l|}{$L_x=N_{\rm{cells},x}\Delta x$ [$\lambda_0/2\pi$]}  & 416   \\ \hline
\multicolumn{1}{|l|}{$\Delta r$ [$\lambda_0/2\pi$]}  & 2 \\ \hline
\multicolumn{1}{|l|}{$N_{\rm{cells},r}$  (half plane)}  & 192  \\ \hline
\multicolumn{1}{|l|}{$L_r=N_{\rm{cells},r}\Delta r$  [$\lambda_0/2\pi$]}  & 384  \\ \hline
\end{tabular}
\caption{Numerical parameters for the LWFA benchmark simulations.}
\label{table:sim_params_benchmark_LWFA}
\end{table}

\begin{figure}[htbp]
\begin{center}
\includegraphics[scale=0.16]{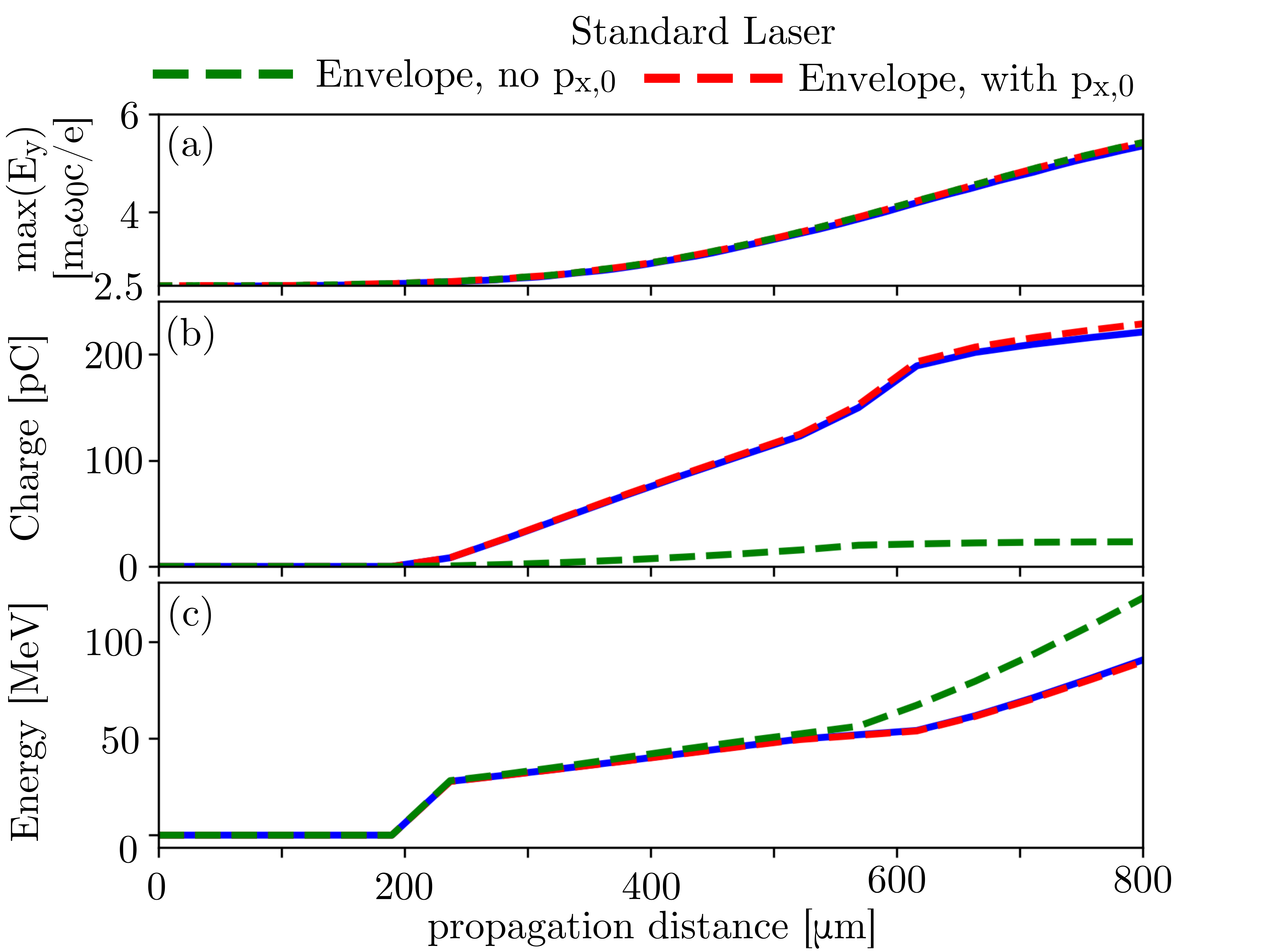}
\caption{\textcolor{black}{(a)} Comparison of the evolution of the laser peak transverse electric field in the LWFA benchmark simulations. For the other two panels, all the electrons created by ionization, present in the moving window and with longitudinal momentum $p_x>50$ $m_ec$, are considered. The electrons injected in the plasma wave bucket behind the laser are a subset of these electrons. \textcolor{black}{(b)} Comparison of the electron total charge evolution; \textcolor{black}{(c)} comparison of the average electron energy.}
\label{fig:a0_charge_energy_evolution_LWFA}
\end{center}
\end{figure}

\begin{figure}[htbp]
\begin{center}
\includegraphics[scale=0.55]{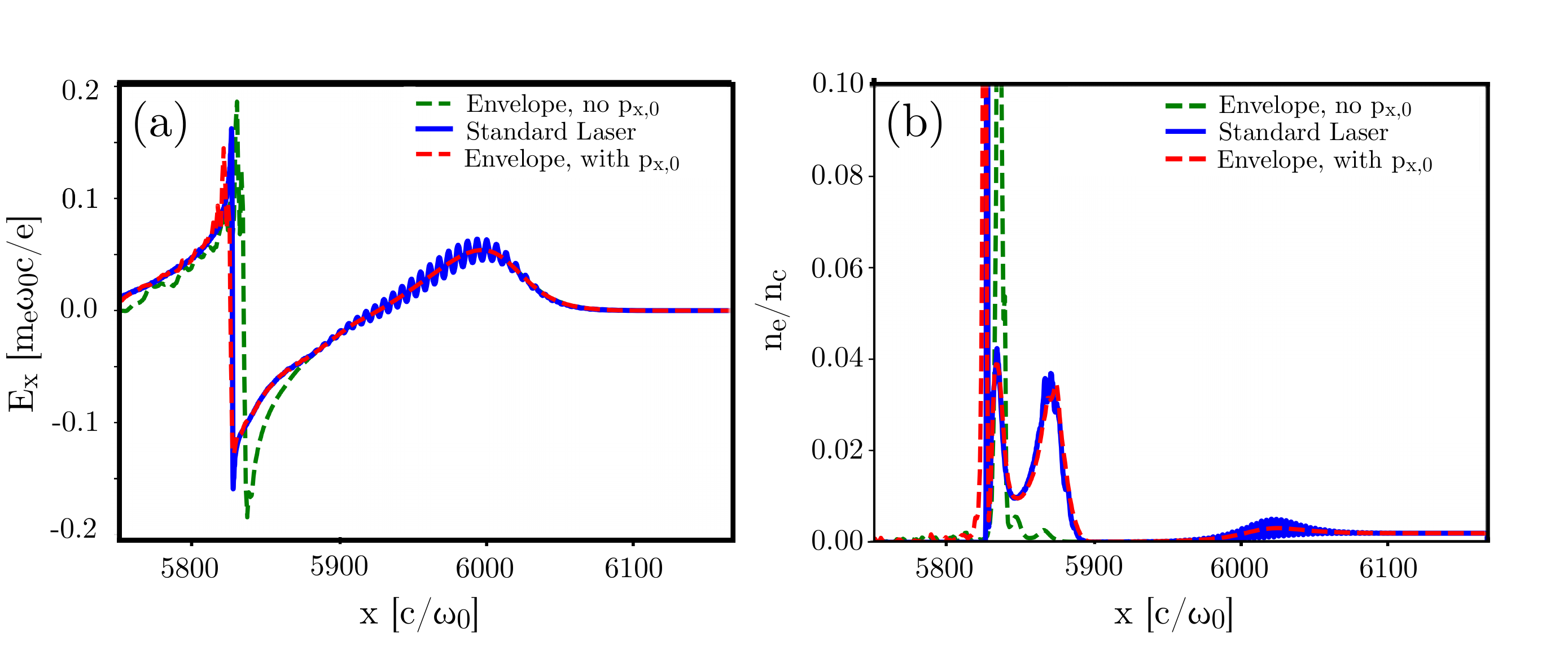}
\caption{Comparison of \textcolor{black}{(a)} the longitudinal electric field and \textcolor{black}{(b) the }electron normalized electron density on the propagation axis after $800$ $\mu$m, computed with the standard laser and envelope simulations in the LWFA benchmark. The maximum value shown for the charge density is $0.01$, to highlight the zone near the injected electron beam.}
\label{fig:Snapshot1D_Ex_Rho_LWFA}
\end{center}
\end{figure}

In Fig. \ref{fig:a0_charge_energy_evolution_LWFA}, the evolution in time of the total charge of the electrons created by ionization and with $p_x>50$ $m_ec$ is compared. A very good agreement between the standard laser and envelope simulation with $p_x$ initialization is found. In the envelope simulation without $p_x$ initialization, the charge is significantly lower ($23$ pC instead of $\approx225$ pC at $800$ $\mu$m for example). As expected, in this simulation the electrons created through ionization do not have an average longitudinal momentum high enough to be trapped in the wakefield behind the laser. In other words, referring for example to Fig. \ref{fig:phase_space_orbits} and the treatment in \cite{Chen2012,FaureCAS}, their initial phase and average momentum in the wakefield lie within the region of untrapped orbits in the phase space, i.e. outside the separatrix curve. Thus, in an envelope simulation, to correctly model these conditions, the averaged initial momentum of the new-born electrons must be properly initialized, otherwise too many electrons would start from a point in the phase space outside the separatrix curve and would not be trapped.
The evolution of the average energy of the same electrons is reported in the bottom panel of Fig. \ref{fig:phase_space_orbits}. A good agreement is found between the standard laser and envelope simulation with $p_x$ initialization. The average accelerating gradient and thus the final energy in the envelope simulation without the $p_x$ initialization are higher. With lower charge trapped in the wakefield bucket  behind the laser, the beamloading effect \cite{Rechatin2009} on the longitudinal electric field in this simulation is significantly lower and a higher $E_x$ field accelerates the macro-particles. To better show this phenomenon, in Fig. \ref{fig:Snapshot1D_Ex_Rho_LWFA} the longitudinal electric field  and the electron charge density on the propagation axis are reported for the three simulations after $800$ $\mu$m of propagation. The beamloading of the longitudinal electric field and the density perturbation in the bunch zone of the standard laser simulation are well reproduced by the envelope simulation with $p_x$ initialization, ensuring a more accurate estimate of the final energy ($\approx90$ MeV), while without the $p_x$ initialization its value is overestimated ($\approx120$ MeV).

In the standard laser and envelope simulation with $p_x$ initialization sudden changes of the charge and energy evolution curve slopes at $\approx550$ $\mu$m and $\approx600$ $\mu$m of propagation occur (see bottom panels of Fig. \ref{fig:a0_charge_energy_evolution_LWFA}). These changes are caused by a complex interplay of self-focusing, injection of a second bunch immediately after the first one (see right panel of Fig. \ref{fig:Snapshot1D_Ex_Rho_LWFA}) and consequent change in beam loading. The study of the charge and energy evolution of this particular set-up is beyond the scope of this work. The important aspect for its scope is that, with the $p_x$ initialization, the envelope simulation can reproduce this behaviour present in the standard laser simulation.

As can be seen in Fig. \ref{fig:a0_charge_energy_evolution_LWFA}, the peak laser electric field value is increasing due to relativistic-self focusing, arriving even at values of $5$ $m_e \omega_0 c/e$. This shows that the accuracy of the ionization procedure with $p_x$ initialization seems robust even at these values of the ionizing field. 

\begin{figure}[htbp]
\begin{center}
\includegraphics[scale=0.6]{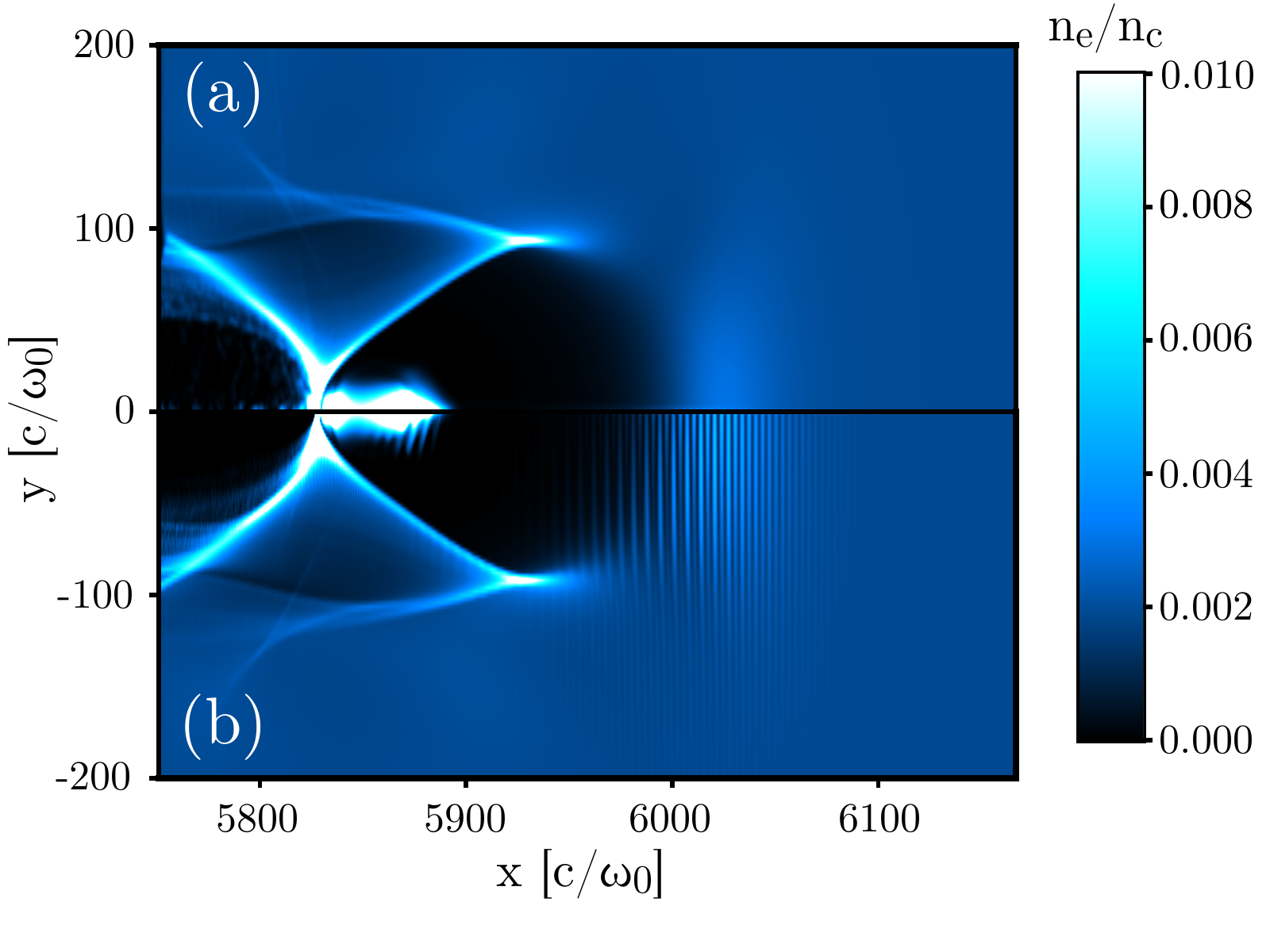}
\caption{Comparison of the electron normalized density on the $xy$ plane after $800$ $\mu$m of propagation computed with \textcolor{black}{(a)} the envelope simulation and \textcolor{black}{(b)} the standard laser simulation in the LWFA benchmark. }
\label{fig:Snapshot_rho_LWFA}
\end{center}
\end{figure}

In Fig. \ref{fig:Snapshot_rho_LWFA}, a snapshot of the electron density after $800$ $\mu$m of propagation is shown, comparing the results of the standard laser and envelope simulation with $p_x$ initialization. Although more accurate comparisons of the electron density should be done in 1D (as in Fig. \ref{fig:Snapshot1D_Ex_Rho_LWFA}, right panel) without the saturation of a colormap in 2D, the shape of the injected bunch appears very similar, apart from a fishbone-like shape in the standard laser simulation. This phenomenon could be caused by the nonlinear mapping in the phase space of the electron discrete injection by laser ionization described in \cite{Xu2016}. Further investigation however is necessary to verify this hypothesis. If the mechanism in \cite{Xu2016} is indeed the  cause of this bunched structure, therefore it cannot be correctly described by an envelope simulation with our model, since the injection does not take place in a discrete way near the peaks of the laser field. However, the bunch parameters do not seem too different between the two simulations (see Table \ref{table:beam_params_LWFA}), hence this phenomenon does not appear to have significant consequences at this distance.

\begin{figure}[htbp]
\begin{center}
\includegraphics[scale=0.15]{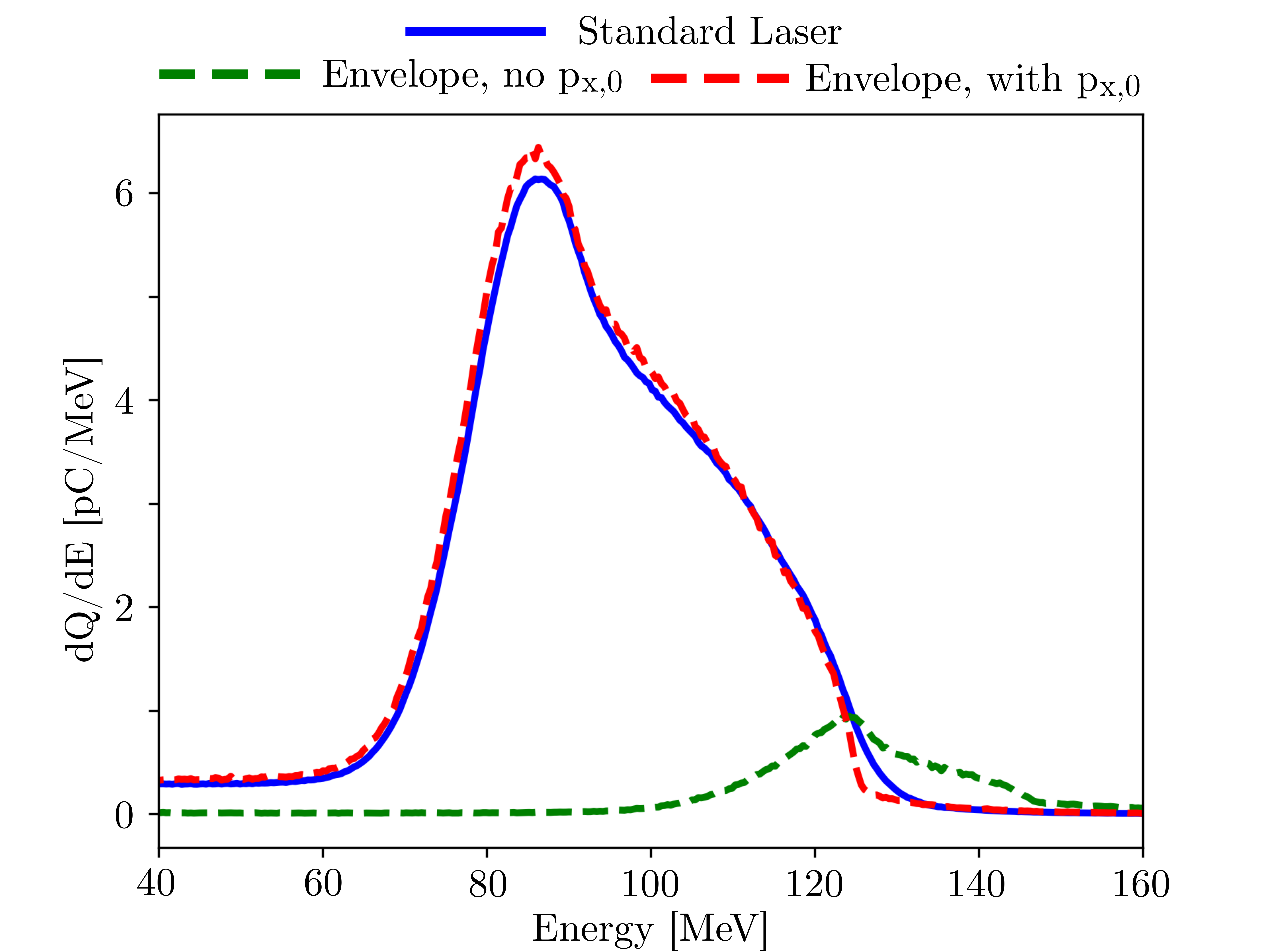}
\caption{Comparison of the energy spectrum of all the electrons created by ionization in the moving window after $800$ $\mu$m of propagation computed with the standard laser and envelope simulations in the LWFA benchmark.}
\label{fig:spectrum_LWFA}
\end{center}
\end{figure}

In Fig. \ref{fig:spectrum_LWFA} the energy spectrum of the electrons created by ionization  with $p_x>50$ $m_e c$ after $800$ $\mu$m of propagation is reported, for the standard laser and envelope simulations. A very good agreement with the laser simulation is found between the laser and envelope simulation with $p_x$ initialization. Without the $p_x$ initialization, as previously discussed, the trapped charge is underestimated and the average energy is overestimated.

Table \ref{table:beam_params_LWFA} summarizes the bunch parameters at the same distance. The bunch is defined as all the electrons within $2.5 \Delta E_{rms} $ of the spectrum peak, where $ \Delta E_{rms} $ is the rms width of the electrons within a FWHM width in the spectrum around the energy peak. Note that the bunch electrons constitute only a subset of the electrons with $p_x> 50\ m_ec$ considered in Fig. \ref{fig:a0_charge_energy_evolution_LWFA}.

\begin{table}[hbtp]
\centering
\renewcommand{\arraystretch}{1}
\begin{tabular}{l|c|c|}
\cline{2-3}
&  \multicolumn{1}{l|}{Standard Laser } & \multicolumn{1}{l|}{Envelope with $p_x$ initialization} \\ \hline
\multicolumn{1}{|l|}{$Q$[$pC$]}  & 175 & 182 \\ \hline
\multicolumn{1}{|l|}{$2\sigma_x$ [$\mu$m]}  & 3.4 & 3.5 \\ \hline
\multicolumn{1}{|l|}{$2\sigma_y$ [$\mu$m]}  & 2.3 & 2.3 \\ \hline
\multicolumn{1}{|l|}{$2\sigma_z$ [$\mu$m]}  & 1.1 & 1.1 \\ \hline
\multicolumn{1}{|l|}{$\varepsilon_{n,y}$ [mm-mrad]}  & 3.9 & 4.0 \\ \hline
\multicolumn{1}{|l|}{$\varepsilon_{n,z}$ [mm-mrad]}  & 1.2 & 1.2 \\ \hline
\multicolumn{1}{|l|}{$E_{avg}$ [MeV]}  & 90.2 & 89.7 \\ \hline
\multicolumn{1}{|l|}{$\sigma_{E}/E$ [rms, \%]}  & 11.91 & 11.93 \\ \hline
\end{tabular}
\caption{Electron beam parameters at $800$ $\mu$m of propagation: charge $Q$, rms sizes $2\sigma_i$ ($i=x,y,z$), normalized emittances $\varepsilon_{n,i}$ ($i=y,z$), mean energy $E$, rms energy spread $\sigma_{E}/E$. First column: beam parameters at the beginning of the simulation. Second and third columns: beam parameters after $800$ $\mu$m of propagation in the standard laser simulation and envelope simulation. The electron beam is defined as all the electrons within $2.5 \Delta E_{rms} $ of the spectrum peak (see Fig. \ref{fig:spectrum_LWFA}), where $ \Delta E_{rms} $ is the rms width of the electrons within a FWHM width in the spectrum around the energy peak.}
\label{table:beam_params_LWFA}
\end{table}

It is important to highlight that he envelope simulation with $p_x$ initialization needed a significantly smaller amount of resources to run (102 cpu-h) compared to the standard laser simulation (9.3 kcpu-h), a factor 91 of difference. This speed-up comes from considering only one azimuthal mode and using larger $\Delta x$ and $\Delta t$ in the envelope simulation.

Thus, with the envelope ionization technique proposed in this work, preliminary envelope LWFA simulations with ionization injection, even with relativistic values for $a_0$, can become affordable with a small cluster. For even quicker simulations, a hybrid fluid-PIC approach \cite{Benedetti2010,MassimoJCP2016,Tomassini2016MatchingSF,Terzani2019} can be envisaged in weakly nonlinear regimes, modeling the immobile ions with macro-particles, the background electrons as a relativistic cold fluid and the electrons created with ionization with macro-particles. In the next section it is shown how the envelope simulation can yield results that are quantitatively accurate enough for preliminary studies with even less resources. 

\section{Effects of reducing the number of particles}\label{LWFAbenchmark_convergence}
In the previous section it was shown that using the same spatial sampling by the macro-particles in the $x-r$ plane and the same number of macro-particles per cell a LWFA envelope simulation with  $p_x$ initialization can yield results that are very similar to those of a standard laser simulation. This agreement was obtained at a propagation distance of $800$ $\mu$m, that is sufficient for a preliminary parametric study, e.g. to design the laser and plasma parameters of an experiment. Since the parameter space to explore is vast, having an estimate of the injected charge and the bunch energy with quick preliminary simulations can greatly speed-up the design process, and this is exactly the purpose of reduced models. More cumbersome simulations with non-reduced models can then investigate further a region of parameters of interest found with a coarse study made with reduced models. In this section it is shown that even degrading the accuracy of the envelope simulation by reducing the number of macro-particles the accuracy of the results remains acceptable, especially considering the reduction of the needed computing resources. The possibility to use only one azimuthal mode and the absence of high frequency oscillations significantly relaxes the sampling requirements in a cylindrical envelope simulation.

The case study for the `degraded' envelope simulations is the same LWFA simulation of the previous section, with the same physical and numerical parameters, except for the distribution of macro-particles per cell. The results of two envelope simulations are reported, with the regular macro-particles distributions $[N_x,\thinspace N_r,\thinspace N_\theta]=[4,\thinspace 2,\thinspace1]$, $[1,\thinspace 1,\thinspace1]$ respectively. Therefore, these `degraded' envelope simulations have respectively 4 times and 32 times less macro-particles per cell than the envelope  simulation of the previous section. In the following Figures and Tables the results are compared to the standard laser simulation of the previous section, which are taken as reference.

\begin{figure}[htbp]
\begin{center}
\includegraphics[scale=0.16]{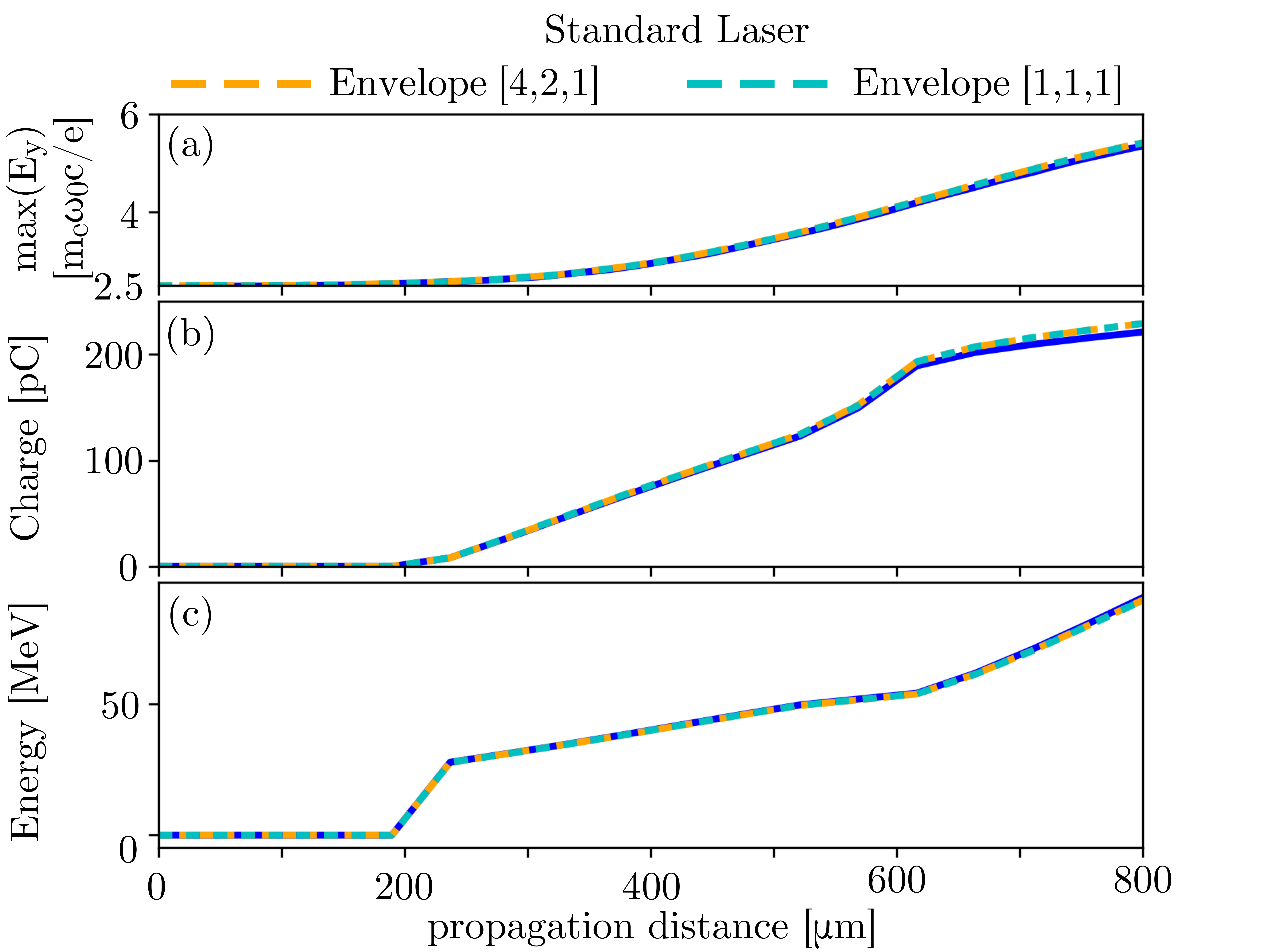}
\caption{\textcolor{black}{(a)} Comparison of the evolution of the laser peak transverse electric field in the LWFA benchmark simulations. For the other two panels, all the electrons created by ionization, present in the moving window and with longitudinal momentum $p_x>50$ $m_ec$, are considered. The electrons injected in the plasma wave bucket behind the laser are a subset of these electrons. \textcolor{black}{(b)} Comparison of the electron total charge evolution;  \textcolor{black}{(c)} comparison of the average electron energy. The results of the standard laser simulation and of the envelope simulations with with $[N_x,\thinspace N_r,\thinspace N_\theta]=[4,\thinspace 2,\thinspace 1],\thinspace[1,\thinspace 1,\thinspace 1]$ are reported. }
\label{fig:a0_charge_energy_evolution_LWFA_reduced_envelope}
\end{center}
\end{figure}

Figure \ref{fig:a0_charge_energy_evolution_LWFA_reduced_envelope} compares the evolution in time of the peak transverse electric field and of the total charge in the moving window computed from the electrons created by ionization with $p_x>50$ $m_ec$ in the standard laser simulation and envelope simulations. The evolution of the average energy of the same electrons is reported as well. Reducing the number  of particles per cell does not seem to significantly influence the integrated charge and average energy of the considered electron population.

\begin{figure}[htbp]
\begin{center}
\includegraphics[scale=0.55]{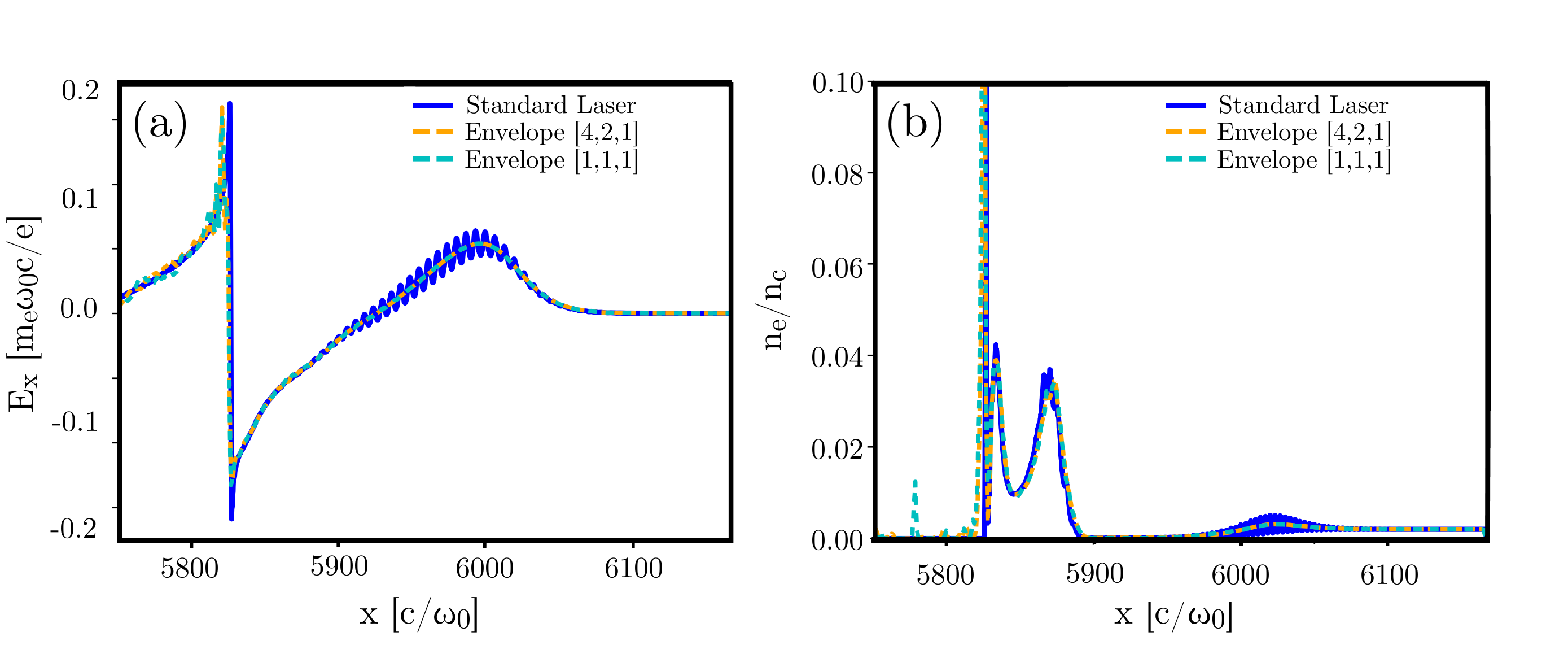}
\caption{Comparison of \textcolor{black}{(a)} the longitudinal electric field and \textcolor{black}{(b) the} electron normalized density on the propagation axis after $800$ $\mu$m, computed with the standard laser and envelope simulations in the LWFA benchmark. The maximum value shown for the charge density is $0.01$, to highlight the zone near the injected electron beam.}
\label{fig:Snapshot1D_Ex_Rho_LWFA_reduced_envelope}
\end{center}
\end{figure}

Figure \ref{fig:Snapshot1D_Ex_Rho_LWFA_reduced_envelope} reports a comparison of the longitudinal electric field and of the electron charge density on axis computed with the standard laser and  envelope simulations. As expected, reducing the number of particles increases the noise in these grid quantities, but the zone near the injected electron beam, where physical phenomena of interes for LWFA occur, displays a high degree of agreement with the standard laser simulation.

\begin{figure}[htbp]
\begin{center}
\includegraphics[scale=0.15]{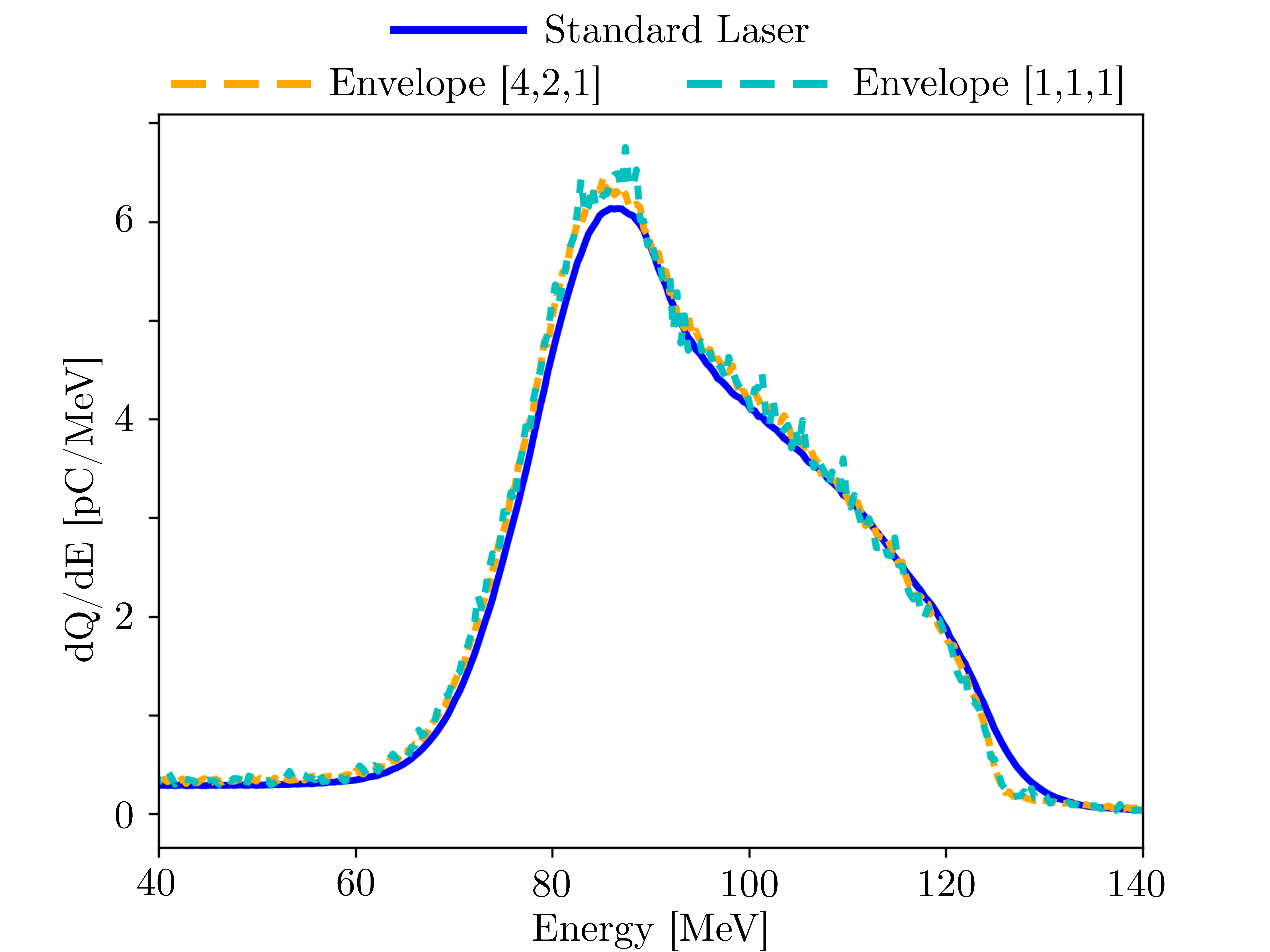}
\caption{Comparison of the energy spectrum of all the electrons created by ionization in the moving window after $800$ $\mu$m of propagation computed with the standard laser and envelope simulations with $[N_x,\thinspace N_r,\thinspace N_\theta]=[4,\thinspace 2,\thinspace 1],\thinspace[1,\thinspace 1,\thinspace 1]$ in the LWFA benchmark.}
\label{fig:spectrum_LWFA_reduced_envelope}
\end{center}
\end{figure}

Figure \ref{fig:spectrum_LWFA_reduced_envelope} compares the spectrum of the electrons obtained with the standard laser simulation and with the envelope simulations. Again, reducing the number of particles increases the level of noise, but the main features of the spectrum are well reproduced even with only one particle per cell. To delve into the details of the accelerated bunch, Table \ref{table:beam_params_LWFA_ppc} compares the beam parameters of the envelope simulations and of the standard laser simulation. Even in the most noisy envelope simulation the beam parameters have a high degree of agreement with the standard laser simulation. The computing time needed for this envelope simulation ($[N_x,\thinspace N_r,\thinspace N_\theta]=[1,\thinspace 1,\thinspace1]$) was 35 minutes, without using MPI or OpenMP. Therefore, with the ionization procedure presented in this work applied to the cylindrical geometry, envelope simulations with results reasonably accurate for preliminary studies can be carried out even from a laptop.

\begin{table}[hbtp]
\centering
\renewcommand{\arraystretch}{1}
\begin{tabular}{l|c|c|c|}
\cline{2-4}
&  \multicolumn{1}{l|}{Standard Laser} & \multicolumn{1}{l|}{Envelope $[N_x,\thinspace N_r,\thinspace N_\theta]=[4,\thinspace 2,\thinspace 1]$} & \multicolumn{1}{l|}{Envelope $[N_x,\thinspace N_r,\thinspace N_\theta]=[1,\thinspace 1,\thinspace 1]$}\\ \hline
\multicolumn{1}{|l|}{$Q$[$pC$]}  & 175 & 182  & 179\\ \hline
\multicolumn{1}{|l|}{$2\sigma_x$ [$\mu$m]}  & 3.4 & 3.5 & 3.5 \\ \hline
\multicolumn{1}{|l|}{$2\sigma_y$ [$\mu$m]}  & 2.3 & 2.3 & 2.4 \\ \hline
\multicolumn{1}{|l|}{$2\sigma_z$ [$\mu$m]}  & 1.1 & 1.1 & 1.2 \\ \hline
\multicolumn{1}{|l|}{$\varepsilon_{n,y}$ [mm-mrad]}  & 3.9 & 4.0 & 4.0\\ \hline
\multicolumn{1}{|l|}{$\varepsilon_{n,z}$ [mm-mrad]}  & 1.2 & 1.1 & 1.2  \\ \hline
\multicolumn{1}{|l|}{$E_{avg}$ [MeV]}  & 90.2 & 89.6 & 89.6 \\ \hline
\multicolumn{1}{|l|}{$\sigma_{E}/E$ [rms, \%]}  & 11.91 & 11.95 & 11.52 \\ \hline
\end{tabular}
\caption{Comparison of standard laser simulation and envelope simulations with $p_x$ initialization and with decreasing number of macro-particles per cell at $800$ $\mu$m of propagation. The Electron beam parameters: charge $Q$, rms sizes $2\sigma_i$ ($i=x,y,z$), normalized emittances $\varepsilon_{n,i}$ ($i=y,z$), mean energy $E$, rms energy spread $\sigma_{E}/E$. First column: beam parameters at the beginning of the simulation. Second and third columns: beam parameters after $800$ $\mu$m of propagation in the standard laser simulation and envelope simulation. The electron beam is defined as all the electrons within $2.5 \Delta E_{rms} $ of the spectrum peak (see Fig. \ref{fig:spectrum_LWFA}), where $ \Delta E_{rms} $ is the rms width of the electrons within a FWHM width in the spectrum around the energy peak.}
\label{table:beam_params_LWFA_ppc}
\end{table}

\section*{Conclusions}
In the context of an existing ionization algorithm for LWFA simulations with an envelope model, an extension of this algorithm was presented, showing a good agreement with standard laser LWFA simulations also for $a_0>1$. This feature proves useful for simulations involving high Z dopant gases like nitrogen where the last ionization levels are accessed with lasers driving highly nonlinear wakefields.
The novel feature of the proposed algorithm is the initialization of the longitudinal momentum  $p_x$ of the new electrons created by ionization, reproducing the initial electron drift in the $x$ direction which becomes significant with high intensity lasers. This extended ionization procedure does  not need an additional finer grid to reproduce the quiver motion of electrons and can be used in Cartesian and cylindrical geometries as well. Two benchmarks have been presented, the ionization of a slab of N$^{5+}$ and a full LWFA simulation with ionization injection with plasma containing N$^{5+}$ as dopant. In the first benchmark it was shown how this $p_x$ initialization allows to accurately reproduce the momenta and evolved positions of the electrons created by ionization, with both linear and circular polarization. In the nonlinear LWFA  benchmark the $p_x$ initialization has been proven to be  essential to accurately compute the total charge trapped in the wakefield bucket behind the laser driver and the other statistical parameters of the trapped electron bunch like average energy, emittance, energy spread, energy spectrum and rms sizes. It was shown that an estimate of these parameters sufficiently accurate for preliminary studies can be obtained even reducing the number of macro-particles per cell to one, case in which the LWFA benchmark can run on a laptop in less than one hour. Considering the degree of agreement obtained for these parameters after $800$ $\mu$m of propagation with significanlty smaller amounts of resources, the proposed extended envelope ionization algorithm can pave the way to quick preliminary studies of LWFA with ionization injection. Future studies should check the effects of the envelope ionization procedure on the accelerated bunches at longer distances of propagation, where also long term 3D effects due to the asymmetry of the linear polarization can become significant.

\appendix
\section{Equations of the envelope model in \Smilei}\label{appendixA}
Under the hypothesis of a laser pulse propagating in the positive $x$ direction, with complex envelope of the transverse component of the vector potential $\mathbf{\tilde{A}}_\perp(\mathbf{x},t)$ slowly varying along the $x$ and transverse directions (e.g a Gaussian laser pulse  with large waist and long duration) compared to the laser wavelength $\lambda_0$, a perturbative treatment can be formulated. This derivation leads to averaged equations which describe some phenomena in the laser-plasma interaction in terms of $\mathbf{\tilde{A}}_\perp$. In many cases of interest for LWFA, this reduced formulation of laser-plasma interaction can accurately describe the relevant physical phenomena that are involved, e.g. self-focusing of the laser pulse, self-injection of electrons and the radiation pressure/ponderomotive force of the laser acting on the particles. The details of this perturbative treatment leading to the following equations can be found in many references, e.g. \cite{Mora1997,Quesnel1998,Cowan2011,Terzani2019}. In this theoretical framework, upon which the envelope simulation model of this work is based, the physical quantities as macro-particles momenta $\mathbf{p}$ can be written as a quickly oscillating part $\mathbf{\hat{p}}$ (denoted with a hat) added to a slowly varying part $\mathbf{\bar{p}}$ (denoted with a bar). The slowly varying positions $\mathbf{\bar{x}}$, momenta $\mathbf{\bar{p}}$ and the so called ponderomotive Lorentz factor $\bar{\gamma}$ of the electrons in an envelope simulation follow the ponderomotive equations of motion \cite{Cowan2011,Terzani2019}:

\begin{equation}\label{ponderomotive_dynamics}
\frac{d\mathbf{\bar{x}}}{dt}=\frac{\mathbf{\bar{p}}}{\bar{\gamma}},\quad \frac{d\mathbf{\bar{p}}}{dt}=-\left(\mathbf{\bar{E}}+ \frac{\mathbf{\bar{p}}}{\bar{\gamma}}\times\mathbf{\bar{B}}\right)-\frac{1}{2\bar{\gamma}}\nabla\Phi,\quad \bar{\gamma} =\sqrt{1+|\mathbf{\bar{p}}|^2+\Phi},
\end{equation}
where $\Phi=|\mathbf{\tilde{A}}_\perp|^2/2$ is the ponderomotive potential.

From d'Alembert's inhomogeneous wave equation for $\mathbf{\hat{A}}_\perp=\Re[\mathbf{\tilde{A}}_\perp e^{i(x-t)}]$, the envelope evolution equation in non-comoving coordinates can be derived \cite{Terzani2019}:
\begin{equation}\label{envelope_equation}
\left[\nabla^2+2i\left(\partial_x+\partial_t\right)-\partial_t^2\right]\mathbf{\tilde{A}}_\perp=\chi\mathbf{\tilde{A}}_\perp,\quad \chi = \sum_{p=1}^{\rm{N_{macro-particles}}} \frac{S(\mathbf{\bar{x}}-\mathbf{\bar{x}}_p)}{\bar{\gamma}_p},
\end{equation}
where $\mathbf{\bar{x}}_p$ is the particle $p$ slowly varying part of the position, $S(\mathbf{\bar{x}})$ the shape function of the particle and $\bar{\gamma}_p$ its ponderomotive Lorentz factor. In \cite{Mora1997,Gordon2000,Huang2006,Cowan2011,Silva2019} a reduced version of Eq. \ref{envelope_equation} is used, but in \Smilei the full form of Eq. \ref{envelope_equation} is solved as in \cite{Terzani2019}.

It is worth noting that, once the susceptibility $\chi$ is known at a given timestep, the envelope equation is linear with respect to $\mathbf{\tilde{A}}_\perp$. Thus at each timestep, instead of a vector equation, only a scalar equation can be solved explicitly for a laser of given polarization, determined by the parameter $\varepsilon$:
\begin{equation}\label{general_polarization}
\mathbf{\tilde{A}}_\perp = \mathbf{e}_y\varepsilon \tilde{A}+i\mathbf{e}_z (1-\varepsilon^2)^{1/2} \tilde{A},
\end{equation}
where $\varepsilon=0,1$ for linear polarization or $\varepsilon=\pm 1/\sqrt{2}$ for circular polarization.
The envelope equation can be solved for a non-zero component of the vector potential at each timestep, provided that the ponderomotive potential is defined as $\Phi=|\mathbf{\tilde{A}}_\perp|^2/2=|\tilde{A}|^2/2$ for all polarizations. Thus, for example the envelope equation could be solved for $\tilde{A}/\sqrt{2}$ in circular polarization and for $\tilde{A}$ in linear polarization (and storing these values would be practical for an easy comparison with a standard laser simulation), but the ponderomotive potential would still be $\Phi=|\mathbf{\tilde{A}}_\perp|^2/2=|\tilde{A}|^2/2$ for both polarizations. This method is used in \Smilei to facilitate the comparison with the laser electric field computed with standard laser simulations. The numerical schemes and the steps used in \Smilei to solve Eqs. \ref{ponderomotive_dynamics}, \ref{envelope_equation} are detailed in \cite{Terzani2019,Massimo2019,Massimo2019cylindrical}.

To easily compare the electric field components between envelope simulations and standard laser simulations, and in the specific context of  this work to correctly compute the total electric field of the laser envelope that might trigger injection (see section \ref{ADK_AC_ioniz_rate}), it is necessary to express the envelopes of the electric field components in terms of the envelope of the transverse vector potential $\mathbf{\tilde{A}}_\perp$. Without loss of generality, here they are derived in case of linear  polarization along the $y$ axis. Since the real field is related to the envelope field through the identity $\hat{A}_y=\Re\left[\tilde{A}_y\thinspace e^{i(x-t)}\right]$, the $y$ component of the electric field can be written, similarly:
\begin{equation}
\hat{E}_y = -\partial_t \hat{A}_y = \Re\left[-\left(\partial_t-i\right)\tilde{A}_y\thinspace e^{i(x-t)}\right]=\Re\left[\tilde{E}_y\thinspace e^{i(x-t)}\right].
\end{equation}
Thus, the envelope of the electric field along $y$ can be defined as
\begin{equation}
\tilde{E}_y = -\left(\partial_t-i\right)\tilde{A}_y.
\end{equation}
For the envelope of the electric field longitudinal component $\tilde{E}_x$, it is useful to change variables, using ($\xi=x-t$, $\tau=t$) $\rightarrow$ ($\partial x=\partial_\xi$, $\partial_t=\partial_\tau-\partial_\xi$). In the context of the envelope perturbative treatment presented in \cite{Mora1997,Quesnel1998,Cowan2011,Terzani2019}, $\partial_\xi\gg\partial_\tau$, thus, using also the Coulomb gauge $\partial_x \hat{A}_x+\partial_y  \hat{A}_y = \partial_\xi \hat{A}_x+\partial_y  \hat{A}_y = 0$, the longitudinal component of the electric field can be found:
\begin{equation}
\hat{E}_x=\Re[\tilde{E}_x e^{i(x-t)}]=-\partial_t \hat{A}_x = \partial_t  \hat{A}_x +\partial_\xi \hat{A}_x \approx \partial_\xi \hat{A}_x = -\partial_y \hat{A}_y=\Re[(-\partial_y\tilde{A}_y )e^{i(x-t)}].
\end{equation}
Thus, the envelope of the longitudinal electric field can be defined as 
\begin{equation}
\tilde{E}_x=-\partial_y \tilde{A}_y.
\end{equation}
Once the envelopes of the  electric field components are computed, the total envelope field defined as  $|\tilde{E}_{\rm{envelope}}| = \sqrt{|\tilde{E}_x|^2+|\tilde{E}_y|^2}$ can be used in  the calculation of the ADK AC ionization rate in Eq. \ref{ADK_AC}.

\section{Initial momentum of the electrons from envelope ionization}\label{appendixB}
The derivation of the initial momentum for an electron obtained from envelope ionization is discussed. The macro-particles representing such electrons are initialized in an envelope PIC simulation with this momentum, which then evolves under the effect of the ponderomotive force and of the averaged electromagnetic fields. The following derivation assumes that the mentioned electrons are extracted when the force due to the laser field, locally a plane wave, constitutes the dominating force acting on them. Indeed, near the peak of the laser pulse the amplitude of the electromagnetic field due to the wake charge separation is much lower than the amplitude of the transverse electromagnetic field of the laser.

Given a test electron in a plane wave, the relations between the electron momentum and the laser vector potential can be analytically found and are briefly reviewed in this Appendix, following the derivations in \cite{Gibbon,Macchi}. For coherence the hat and bar notation of the previous Appendix for the quickly and slowly oscillating quantities are maintained. The ansatz of the following derivation is a plane wave propagating along the positive $x$ direction, i.e. described by a vector potential with zero component on the $x$ direction and form $\mathbf{A}_{\perp}(x-t)$. From this laser vector potential, the laser transverse electromagnetic fields can be expressed as $\mathbf{E}_{\perp}=-\partial_t \mathbf{A}_{\perp}$, $\mathbf{B}_{\perp}=\mathbf{e}_x\times\partial_x \mathbf{A}_\perp$. The momentum evolution equation for an electron in this plane wave can be rewritten as
\begin{equation}\label{transv_momentum_plane_wave}
\frac{d\mathbf{p}_{\perp}}{dt}=-\mathbf{E}_\perp-\left(\mathbf{v}\times\mathbf{B}_\perp\right)_\perp=(\partial_t+v_x\partial_x)\mathbf{A}_{\perp}=\frac{d\mathbf{A}_{\perp}}{dt},
\end{equation}
which implies $\mathbf{p}_\perp - \mathbf{A}_\perp = \rm{constant}$. In LWFA the atom/ion from which the electrons are extracted through tunneling ionization can be considered at rest, hence for conservation of momentum the extracted electrons can be considered at rest ($p_{x,t_{\rm{ioniz}}}=|\mathbf{p}_{\perp,t_{\rm{ioniz}}}|=0$). Therefore, from Eq. \ref{transv_momentum_plane_wave} it can be inferred that:
\begin{equation}\label{transv_momentum_plane_wave2}
\mathbf{p}_{\perp}-\mathbf{A}_{\perp}=\mathbf{p}_{\perp,t_{\rm{ioniz}}}-\mathbf{A}_{\perp,t_{\rm{ioniz}}}=-\mathbf{A}_{\perp,t_{\rm{ioniz}}}\quad\rightarrow\quad \mathbf{p}_{\perp}=\mathbf{A}_{\perp}-\mathbf{A}_{\perp,t_{\rm{ioniz}}},
\end{equation}
where $\mathbf{A}_{\perp,t_{\rm{ioniz}}}$ is the transverse vector potential acting on the electron initially, at the ionization time $t_{\rm{ioniz}}$. 
Similarly, the equation for the variation of the electron longitudinal momentum and energy can be rewritten:
\begin{eqnarray}
\frac{dp_x}{dt} &=& - \left(\mathbf{v}\times\mathbf{B}_\perp\right)_x=-\left(v_y\partial_x A_z+v_z\partial_xA_y\right)\\
\frac{d\gamma}{dt} &=& - \left(\mathbf{v}\cdot\mathbf{E}_\perp\right)=\left(v_y\partial_t A_y+v_z\partial_tA_z\right)
\end{eqnarray}
Subtraction of the previous Equations and the use of $(\partial_t-\partial_x)\mathbf{A}_{\perp}=0$ yields $\frac{d(p_x-\gamma)}{dt}=0$, implying $p_x-\gamma=\rm{constant}$. Coherently with the hypothesis of electron initially at rest ($p_{x,t_{\rm{ioniz}}}=0$, $\gamma_{t_{\rm{ioniz}}}=1$), and using the identity $\gamma^2=1+p_x^2+|\mathbf{p}_{\perp}|^2$, a quadratic relation between the longitudinal momentum and the vector potential amplitude can be found: 
\begin{equation}\label{longitudinal_momentum_plane_wave}
p_x =\frac{|\mathbf{p}_\perp|^2}{2}= \frac{1}{2}|\mathbf{A}_{\perp}-\mathbf{A}_{\perp,t_{\rm{ioniz}}}|^2.
\end{equation}

An envelope model describes the dynamics of the macro-particles using their averaged positions and momenta, thus an average over the optical cycles is then computed on the momentum components represented by Eqs. \ref{transv_momentum_plane_wave2}, \ref{longitudinal_momentum_plane_wave}:
\begin{eqnarray}
\mathbf{\bar{p}}_\perp&=& \overline{\left(\mathbf{A}_{\perp}-\mathbf{A}_{\perp,t_{\rm{ioniz}}}\right)}=-\mathbf{A}_{\perp,t_{\rm{ioniz}}},\label{averaged_transv_momentum}\\
\bar{p}_x&=&\frac{1}{2}\overline{\left(|\mathbf{A}_{\perp}|^2+|\mathbf{A}_{\perp,t_{\rm{ioniz}}}|^2-2\thinspace\mathbf{A}_{\perp}\cdot\mathbf{A}_{\perp,t_{\rm{ioniz}}}\right)}=\frac{1}{2}\overline{\left(|\mathbf{A}_{\perp}|^2+|\mathbf{A}_{\perp,t_{\rm{ioniz}}}|^2\right)}=\frac{1}{2}\overline{|\mathbf{A}_{\perp}|^2}+\frac{1}{2}|\mathbf{A}_{\perp,t_{\rm{ioniz}}}|^2.\label{averaged_long_momentum}
\end{eqnarray}
To derive the last two identities the constancy of $\mathbf{A}_{\perp,t_{\rm{ioniz}}}$ for a given electron over the optical cycles was used, and a plane wave of the form $\mathbf{A}_\perp=\mathbf{\hat{A}}_\perp=\Re\left[\mathbf{\tilde{A}}_\perp\thinspace e^{i(x-t)}\right]$ with $\mathbf{\tilde{A}}_\perp$ described by Eq. \ref{general_polarization} was assumed.

Using the same plane wave definition, from Eqs. \ref{averaged_transv_momentum}, \ref{averaged_long_momentum} the initial momentum conditions for an electron created through envelope ionization described in sections \ref{p_perp_initialization}, \ref{px_initialization} can be motivated.

In linear polarization, the rms thermal spread $\sigma_{p_\perp}$ of the transverse momentum is due to the extraction of electrons at different phases of the laser, which yield different $\mathbf{A}_{\perp,t_{\rm{ioniz}}}$. This thermal spread is quantified by Eq. \ref{transverse_momentum_initialization}, derived in \cite{Schroeder2014}. Indeed, Eq. \ref{transverse_momentum_initialization} is derived averaging the square of $\mathbf{\bar{p}}_\perp$ (which is obtained through an average over the optical cycles denoted with the bar) over the possible ionization phases, an operation which will be denoted as $<|\bar{\mathbf{p}}_\perp|^2>=<|\bar{\mathbf{p}}_\perp|^2>_{\rm{ioniz\thinspace phases}}$. With this notation, $\sigma_{p_\perp}=(<|\bar{\mathbf{p}}_\perp|^2>)^{1/2}=(<|\mathbf{A}_{\perp,t_{\rm{ioniz}}}|^2>)^{1/2}=\Delta |\tilde{A}|$ (Eq. 9 of \cite{Schroeder2014}), where $\Delta$ is defined in Eq. \ref{eqref:Delta} and \cite{Schroeder2014}. In an envelope simulation, the typical resolution does not allow to resolve the fast phase oscillations of $\mathbf{A}_{\perp,t_{\rm{ioniz}}}$ in linear polarization. Thus, to recreate this rms thermal spread, the initial transverse momentum of electrons in the polarization direction is initialized with a number $p_{pol,\thinspace 0}$ drawn from a centered Gaussian distribution with the mentioned rms spread $\sigma_{p_\perp}$, i.e. $N(0,\sigma_{p_\perp})$.  For circular polarization, the amplitude of the wave does not change with the phase oscillations, thus the value $|\mathbf{A}_{\perp,t_{\rm{ioniz}}}|=|\tilde{A}|/\sqrt{2}$ can be used for the initial electron transverse momentum, randomly assigning the transverse direction.

The quadratic dependence of the longitudinal momentum with respect to the transverse momentum, which is an oscillating function, suggests that the initial longitudinal momentum assigned to the electrons created by envelope ionization $\bar{p}_{x,0}$ should be a non-negative value.
For circular polarization, the plane wave definition that was chosen gives, for Eq. \ref{averaged_long_momentum}, $\bar{p}_{x,0}=|\tilde{A}|^2/2$. From Eq. \ref{averaged_long_momentum} it can be inferred that for linear polarization, the average $<\bar{p}_x>$ over the possible ionization phases of electrons subject to a vector potential with envelope $\tilde{A}$ is dominated by $|\tilde{A}|^2/4$, i.e. $(1/2)\overline{|\mathbf{A}_\perp|^2}$ with $|\mathbf{A}_\perp|=|\tilde{A}|\sin(x-t)$. This represents the first term of the last identity of Eq. \ref{averaged_long_momentum}. Indeed, the average over the ionization phases of the second  term gives $(1/2)\sigma^2_{p_\perp}$, which scales as $\frac{1}{2}[(1.5)^{1/2}|\tilde{A}|^{3/2}(2I_p)^{-3/4}]^2$. For $|\tilde{A}|=2.5$ and the ionization of the sixth level of nitrogen ($I_p=40.6$), its ratio over the first term is $0.01$ . To assign an initial $p_{x,0}$ to an electron created by envelope ionization in linear polarization, in the procedure presented in this work this second small term is neglected. Instead, coherently with the identity $\bar{p}_x=(1/2)|\bar{\mathbf{p}}_{\perp}|^2$, it is reasonable to assign a greater longitudinal momentum to electrons with a greater transverse momentum, thus a term $(1/2)p^2_{pol,\thinspace 0}$ was chosen to be added to $|\tilde{A}|^2/4$ in the initial longitudinal momentum. This choice, although heuristic, also allows to give a stochastic spread in the initial longitudinal momentum of the new electrons, for which no analytical results about its distribution are available at the present moment to the authors' knowledge. Further developments of this procedure include analytical calculations that will allow to initialize the longitudinal momenta with a more formally derived stochastic component.

To resume, dropping the bar notation, in the envelope ionization procedure of this work the magnitude of the initial transverse momentum of the new electrons $|\mathbf{p_{\perp,0}}|$, subject to an envelope $\tilde{A}$, is assigned as in \cite{Tomassini2017}:
\begin{equation}
|\mathbf{p_{\perp,0}}|= \left\{
\begin{array}{ll}
      p_{pol,\thinspace 0} \leftarrow N(0,\sigma_{p_\perp}) & \text{for linear polarization} \\
      \quad|\tilde{A}|/\sqrt{2} & \text{for circular polarization}. \\
\end{array} 
\right. 
\end{equation}
The direction of the initial momentum is the polarization direction for linear polarization and randomly assigned between $0$ and $2\pi$ for circular polarization.

The initial longitudinal momentum of the new electron is assigned with the following choice:
\begin{equation}
p_{x,0}= \left\{
\begin{array}{ll}
      |\tilde{A}|^2/4+p_{pol,0}^2/2 & \text{for linear polarization} \\
      |\tilde{A}|^2/2 & \text{for circular polarization}. \\
\end{array} 
\right. 
\end{equation}

\section*{Acknowledgments}
F. Massimo was supported by P2IO LabEx (ANR-10-LABX-0038) in the framework “Investissements d’Avenir” (ANR-11-IDEX-0003-01) managed by the Agence Nationale de la Recherche (ANR, France) as well as by Laboratoire de physique des deux infinis Irène Joliot-Curie (IJCLab IN3P3-PALLAS project).
The authors are grateful to Kevin Cassou and the IJCLab team for fruitful discussions. 

Computer time has been granted by GENCI (project 2019-A0050510062) and by École polytechnique through the LLR-LSI project.
The authors are grateful to the TGCC and CINES engineers for their support. The authors thank the engineers of the LLR HPC clusters for resources and help. 

The authors are grateful to D. Terzani and Gilles Maynard for fruitful discussions.

\renewcommand{\baselinestretch}{1.0}

\bibliographystyle{unsrt}
\bibliography{Bibliography}

\end{document}